\documentclass[12pt, letterpaper]{article}
\usepackage{graphicx}
\usepackage{multirow}
\usepackage{amsmath,amssymb,amsfonts}
\usepackage{amsthm}
\usepackage{mathrsfs}
\usepackage[title]{appendix}
\usepackage{makecell}
\usepackage{xcolor}
\usepackage{booktabs}
\usepackage{listings}
\usepackage{array}
\usepackage{siunitx}
\usepackage{lineno}
\usepackage{tabularx}
\usepackage[citestyle=authoryear,natbib=true,style=apa]{biblatex}
\DeclareLanguageMapping{american}{american-apa}

\usepackage{geometry}
\begin{document}

\def\scititle{
Distilling human mobility models with \\ symbolic regression
}

\title{\bfseries \boldmath \scititle}
\author{
	Hao Guo$^{1\dagger}$,
	Weiyu Zhang$^{1\dagger}$,
	Junjie Yang$^{1}$,
    Yuanqiao Hou$^{1}$,
    Lei Dong$^{1\ast}$,
    Yu Liu$^{1}$\and 
\normalsize $^{1}$Institute of Remote Sensing and Geographical Information Systems, \and 
\normalsize School of Earth and Space Sciences, Peking University, Beijing, China.\and
\normalsize $^\ast$Corresponding author. Email: leidong@pku.edu.cn\and
\normalsize $^\dagger$These authors contributed equally to this work.
}
\date{}
\maketitle

\begin{abstract}
Human mobility is a fundamental aspect of social behavior, with broad applications in transportation, urban planning, and epidemic modeling. 
Represented by the gravity model and the radiation model, established analytical  models for mobility phenomena are often  discovered by analogy to physical processes. Such discoveries can be challenging and rely on intuition, while the potential of emerging social observation data in model discovery is largely unexploited. Here, we propose a systematic approach that leverages symbolic regression to automatically discover interpretable models from human mobility data. Our approach finds several well-known formulas, such as the distance decay effect and classical gravity models, as well as previously unknown ones, such as an exponential-power-law decay that can be explained by the maximum entropy principle. By relaxing the constraints on the complexity of model expressions, we further show how key variables of human mobility are progressively incorporated into the model, making this framework a powerful tool for revealing the underlying mathematical structures of complex social phenomena directly from observational data.
\end{abstract}


\normalsize
\section{Introduction}

A deeper understanding of population movement and its underlying mechanisms is crucial for informed urban planning \citep{AHS24,WSW21}, transportation system management \citep{CLG16}, and epidemic control strategies \citep{JLY20,SBB23}. Historically, the study of human mobility has evolved from qualitative description of key factors  \citep{Rave85} to quantitative analyses, such as the characterization of the distance decay effect \citep{Lill89}, culminating in formal mathematical models of population flows. The most prominent of these is the gravity model, proposed in the early 20th century \citep{Stew41,Zipf46}. Analogous to Newton's law of gravitation, the gravity model posits that human mobility flows are proportional to the population of interacting locations and inversely related to the distance between them \citep{RoTh04, And11}. 

\par Despite the enduring influence of the class of gravity models \citep{Wil70}, progress in developing analytical models for population flows has been intermittent. Notable contributions include the intervening opportunity model \citep{Stou40,Sch59}, the radiation model \citep{SGM12}, and models incorporating preferential return \citep{SKW10,BdLNE15,SDOK21}. A major challenge within this physics-informed paradigm has been the lack of analytical frameworks derived from first principles. Unlike the physical sciences where models are often grounded in fundamental laws, models describing social phenomena, such as mobility behavior, are predominantly empirical, relying on intuition and interactive exploration to identify appropriate model forms. The inherent complexity and heterogeneity of human movement further compound this challenge. 

On the other hand, there has been rapid development of machine-learning models for human mobility, especially those based on artificial neural networks \citep{SBL21, LHF25} and graph neural networks \citep{LMX20, YGZ21, LZW24}. These models typically incorporate multiple features (e.g., points of interest, road density, land use) to achieve better prediction accuracy. Yet their contribution to understanding human mobility has been limited due to their black-box nature.

The recent availability of high-resolution mobility data from sources such as call detail records (CDR), real-time smartphone tracking, and social media check-ins \citep{BBG18,WKS19,PMS23} has offered unprecedented opportunities to investigate human mobility. These datasets provide valuable insights into both routine commuting patterns and more sporadic travel behaviors. However, the high dimensionality and inherent noise in these datasets pose substantial challenges for model identification using traditional, intuition-based approaches \citep{Cran23}. This necessitates automated methods capable of leveraging large observational datasets to accelerate the discovery of accurate and interpretable models of human mobility.

The endeavor to build symbolic mobility models automatically dates back to \citet{Open88} and \citet{Dip98}. Nevertheless, the potential of this paradigm was not fully realized due to inefficient mobility data and limitations in model optimization methods at that time. Without proper regularization of model complexity, the discovered expressions are lengthy and difficult to interpret. Nowadays, Symbolic Regression (SR) has emerged as a powerful technique to uncover analytical expressions directly from data \citep{MaCa24}, with applications in diverse scientific disciplines \citep{RPSP20,WSZ20,WTVN23,LZW24,VeDo22,LLS24}. SR learns both the function structure and the associated parameters, relieving the need to specify the function form \emph{a priori} \citep{LCOB21}. This flexibility enables the discovery of novel scientific laws and provides insights into the underlying mechanisms of complex phenomena \citep{CBB20}. Moreover, SR has shown the ability to produce more generalizable models than other machine learning techniques, such as neural networks, due to its inherent preference for simpler, more compact expressions \citep{VNAAG21, LJC23, CTDM25}. 

Here, we propose a Symbolic Regression-based framework for discovering models of human mobility. We apply this approach to mobility data from three countries - China, the UK, and the US - examining both commuting and general mobility flows. Our results show that the exponential decay gravity model consistently emerges across countries and spatial scales, suggesting its optimality in describing human mobility under certain model complexities  (i.e. expression length). Beyond recovering known models, SR uncovers novel extensions, including a previously unknown exponential power-law distance decay, which can be interpreted through the maximum entropy principle \citep{Wil70}. Furthermore, by analyzing symbolic models derived from flow data grouped by geographic regions, we reveal notable heterogeneity in model forms between intra- and inter-region flows, providing new insights into the diverse modes of human movement that are often masked in global models. Finally, using synthetic flow data, we validate the robustness of SR to discover various mobility models from noisy observations. 

\section{Research design}

\subsection{The analytical framework}

\begin{figure}[htbp]
\centering
\includegraphics[width=\linewidth]{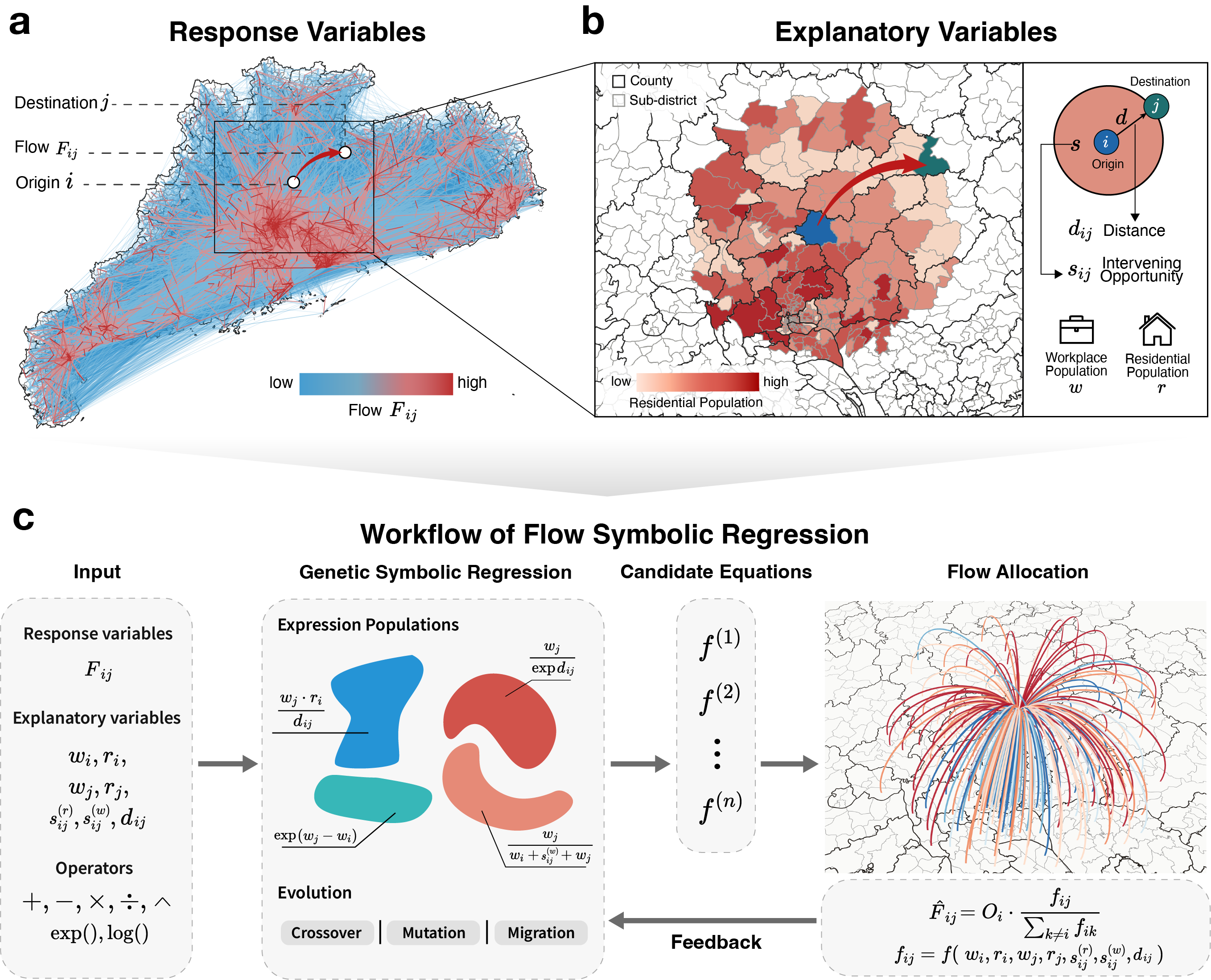}
\caption{\fontsize{11}{15}\linespread{1}\selectfont{\textbf{The analytical framework of mobility model distillation.}} (\textbf{a}) Mobility flows in Guangdong, China. The flow volume $F_{ij}$ from origin $i$ to destination $j$ is the response variable. (\textbf{b}) The explanatory variables include the workplace population $w_i, w_j$, the residential population $r_i, r_j$, geographic distance $d_{ij}$, and intervening opportunities $s^{(w)}_{ij},s^{(r)}_{ij}$, calculated with workplace and residential population, respectively. (\textbf{c}) The overall workflow to automatically distill models from mobility data. Considering seven common operators, we use a genetic-programming-based SR program to search for appropriate model forms. In each iteration, the SR program generates models for allocation weight function $f$. For each origin $i$, the total outflow $O_i$ is allocated to each destination based on the corresponding allocation weights $f_{ij}$. The MSE between the predicted flow $\hat{F}_{ij}$ and the actual flow $F_{ij}$ is then calculated and fed back into the SR program for expression optimization.}
\label{fig:framework}
\end{figure}

Figure~\ref{fig:framework} presents the analytical framework. To distill mathematical models of human mobility, we first compiled four large-scale datasets from three countries: 2020 cellphone data from Guangdong Province, China; 2019 cellphone data from Beijing-Tianjin-Hebei urban agglomeration, China; 2011 Census data from England; and 2011-2015 American Community Survey (ACS) data from the US (see Section \ref{sec:data} for details). These datasets encompass commuting flows (both cellphone and survey data) and general mobility patterns (cellphone data) across various spatial resolutions and geographical areas, thereby enabling a comprehensive evaluation of the robustness of potential mobility models.

Given a dataset with spatial units indexed by $i=1,2,\dots,n$, we consider four types of explanatory variables: workplace population $w_i$, residential population $r_i$, geographical distance $d_{ij}$, and intervening opportunities $s_{ij}$ (Fig.~\ref{fig:framework}b). Here, intervening opportunities represent the number of competing destinations that are closer to the origin \citep{BBG18}. In line with the literature, we use  population as a proxy of competing destinations \citep{SGM12}. As we consider both residential and workplace population, we calculate intervening opportunities based on each of them:
\begin{align}
    s_{ij}^{(r)} &= \sum_{k:0<d_{ik}<d_{ij}} r_k \\ 
    s_{ij}^{(w)} &= \sum_{k:0<d_{ik}<d_{ij}} w_k.
\end{align}
Besides variables, we consider five basic binary operators (Fig.~\ref{fig:framework}c): addition ($+$), subtraction ($-$), multiplication ($\times$), division ($\div$), power (\^{}); and two unary operators: exponentiation ($\exp$), logarithm ($\ln$). 

\par
\begin{figure}
\centering
\includegraphics[width = 0.6\linewidth]{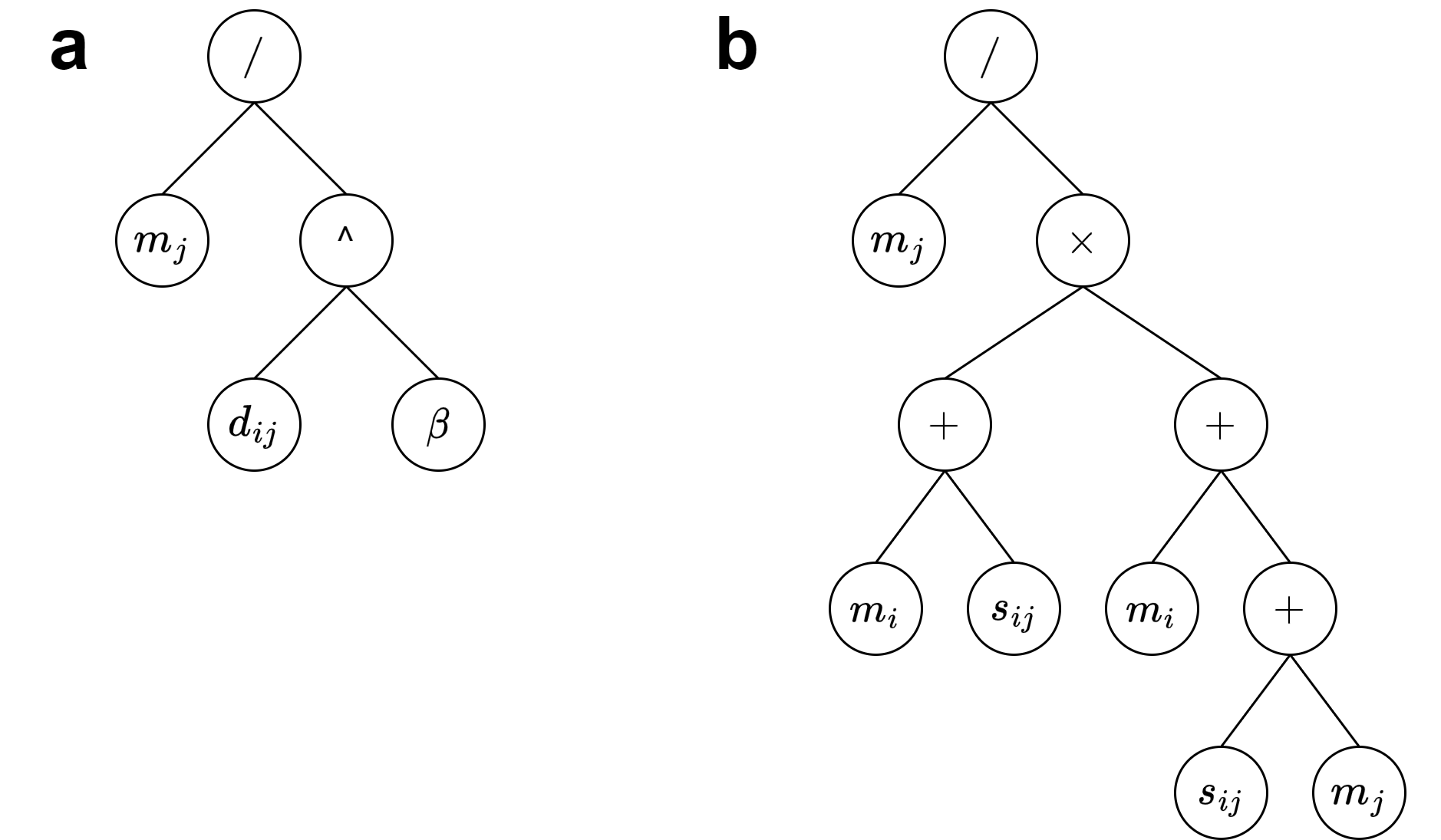} 
\caption{\fontsize{11}{15}\linespread{1}\selectfont{\textbf{Examples for calculating the model complexity with binary expression trees.}} (\textbf{a}) Under the allocation weight setting (see Section 2.3), the gravity model with power-law decay $f_{ij}=m_j/d_{ij}^\beta$ is expressed as a tree with 5 nodes, hence has a complexity of 5. (\textbf{b}) The radiation model $f_{ij}=m_j/(m_i+s_{ij})(m_i+s_{ij}+m_j)$ has a complexity of 11. }
\label{fig:tree}
\end{figure}

 The goal of symbolic regression is to identify the optimal model form by simultaneously optimizing the accuracy and complexity. Specifically, accuracy is measured by the mean squared error (MSE) of the flow volumes over the entire dataset, while complexity is quantified by the length of the expression, defined as the total number of variables, operators, and constants, with repeated symbols counted separately \citep{Cran23}. Each constant, variable, and binary operator is assigned a length of 1, while $\exp$ and $\ln$ are assigned a length of 2 due to the implicit presence of the constant $e$. When the expression is represented as a full binary tree, model complexity corresponds to the number of nodes, which must always be an odd integer (see Fig.~\ref{fig:tree} for examples). 

To identify the optimal model form, we apply a genetic programming-based SR algorithm (Fig.~\ref{fig:framework}c). This algorithm evolves a population of expressions represented as trees. In analogy to natural selection, expressions with higher fitness values (calculated based on accuracy and complexity) have a higher probability of producing offspring through crossover and mutation. The algorithm generates a Pareto frontier of optimal expressions, with each expression being Pareto-optimal if no improvement in one objective (accuracy or complexity) can be made without sacrificing the other. Detailed SR settings are given in Section \ref{sec:srsetting}.

We use the root mean squared error (RMSE) and common part of commuters (CPC) \citep{LHG12} to measure the discrepancy between the model-predicted flows and real flows. CPC is a common metric for models of origin-destination flows \citep{LBR16, SBL21}. It is defined as:
\begin{equation}
    \text{CPC}=\frac{2\sum_{i,j}\min \{F_{ij}, \hat{F}_{ij} \} }{\sum_{i,j} F_{ij} + \sum_{i,j} \hat{F}_{ij}},
\end{equation}
where $F_{ij}, \hat{F}_{ij}$ are the real and predicted flow volume from place $i$ to $j$, respectively. The range of CPC is $[0,1]$, and higher value indicates a better match between real and predicted flows. 

We compare the SR-derived models with six existing mobility models (Fig.\ref{fig:srmain}e): Zipf's simple gravity model (GMZipf, \cite{Zipf46}); gravity model with power-law/exponential decay (GMPow/GMExp, \cite{Wil70});  opportunity priority selection model (OPS, \cite{LiYa19}); radiation model (RM, \cite{SGM12}); Schneider's intervening opportunity model (IO,  \cite{Sch59}). These models are also calibrated to minimize MSE. For commuting datasets, we use workplace population (and intervening opportunities based on workplace population) in the existing models, as using residential population yields inferior results.

\subsection{Data} 
\label{sec:data}

A general description of the flow datasets is given in Table~\ref{tab:dataset}. The mobility flows are visualized in Supplementary Figs. S2-S6. 

The Guangdong dataset is provided by China Mobile Communications Corporation, the largest telecommunications operator in China. The raw dataset comprises individual movement trajectories of 5 million users in November 2020, with a spatial resolution of 500\,m $\times$ 500\,m. After filtering out incorrect or duplicate trajectory points, we identify stay points as trajectory points with a duration of at least 30\,min. The mobility flows are movements between stay points, aggregated at the sub-district and county levels based on the grid cells’ center locations. 

For the commuting flow dataset, an individual’s home is defined as the location where he stayed the longest between 12:00 AM and 7:00 AM, and the workplace is the location where he stayed the longest between 9:00 AM and 6:00 PM. Commuting trips are identified as movements from the home location to the workplace in the morning (before 12:00 PM). Subsequently, the commuting movements are aggregated similarly to the mobility flow dataset. The residential / workplace populations are derived based on the detected home and workplace locations.

The Beijing-Tianjin-Hebei mobility data is provided by China Unicom, which contains inter-county mobility from November 4 to 10, 2019. Similarly, the mobility flows are aggregated movements between stay points with durations $\ge$ 30\,min. Registered population data in 2019 are collected from official statistical yearbooks. 

The commuting datasets in England and the Contiguous US record the location of usual residence and workplace of employed residents, aggregated to merged local authority district (MLAD) and county levels, respectively. The sources of residential and workplace population data are the same as the flow data (2011 Census for England, and 2011-2015 ACS 5-year estimate for US). 

For all datasets above, we only consider flows between spatial units, so flows with the same origin and destination are removed. Moreover, we use the population data at finer scale (if available) to calculate the intervening opportunity. For example, to calculate the intervening opportunities from county $i$ to $j$ in Guangdong, we add up the population of all sub-districts $k$ outside $i$ and $j$ where $d_{ik}<d_{ij}$. Similarly, population at middle-layer super output area (MSOA) level is used for the England dataset. 

\begin{table}
\centering
\caption{\fontsize{11}{15}\linespread{1}\selectfont{\textbf{Description of the real flow datasets.} Area and population are the median values among all spatial units. An observation refers to an origin-destination pair. BTH is short for Beijing-Tianjin-Hebei urban agglomeration, China.}}

\renewcommand{\arraystretch}{1.0} 
\begin{tabular}{ccccccc}
\hline
Region & Granularity & Type & \#Units & \#Observations & Area ($\text{km}^2$)& Population \\
\hline
\multirow{2}{*}{Guangdong} & \multirow{2}{*}{sub-district} & mobility & \multirow{2}{*}{2,100} & 705,811 &  \multirow{2}{*}{60} & \multirow{2}{*}{36,962} \\
&& commuting &&28,392 & & \\
\hline
\multirow{2}{*}{Guangdong} & \multirow{2}{*}{county} & mobility & \multirow{2}{*}{124} & 14,251 & \multirow{2}{*}{1,256} & \multirow{2}{*}{725,391} \\
& & commuting & & 1,150 & & \\
\hline
England & MLAD & commuting & 324 & 57,591 & 212 & 126,430 \\
\hline
US & county & commuting & 3,108 & 131,391 & 1,616 & 25,991 \\
\hline
BTH & county & mobility &  200 & 38,813 & 756 & 440,000 \\
\hline 
\end{tabular}
\label{tab:dataset}
\end{table}

\subsection{Symbolic Regression} 
\label{sec:srsetting}

Rather than directly modeling the origin-destination flow $F_{ij}$ between places $i$ and $j$, we model the allocation weight $f_{ij}$ with symbolic regression. This weight is proportional to the probability that an individual starting from place $i$ will choose destination $j$ (Fig.~\ref{fig:framework}c). Let $O_i$ denote the total outflow from place $i$, $F_{ij}$ is calculated from $\{f_{ik}\}_{k\neq i}$ as follows: 
\begin{equation}
    \hat{F}_{ij}= O_i \frac{f_{ij}}{\sum_{k\neq i} f_{ik} }.
\label{eq:normalize}
\end{equation}
This approach reduces the complexity of the target models. For example, the power-law decay gravity model
\begin{equation}
F_{ij} = k m_i m_j/d_{ij}^{\beta}
\end{equation}
has a complexity of 9, while its allocation weight form 
\begin{equation}
f_{ij} = m_j/d_{ij}^\beta
\end{equation} 
has a complexity of 5, where $m_i$, $m_j$ are the populations of the origin $i$ and destination $j$, respectively. The search space of SR grows exponentially with expression length, which can largely impact the performance of SR to find the optimal function form. Our allocation weight approach reduces the computation burden of SR, enhancing its stability and efficiency. 

Based on the allocation weight approach, our SR problem can be formulated as a minimisation problem to find
\begin{equation}
    f^* \in \mathop{\arg\min}\limits_{f \in \mathcal{F}} \frac{1}{n(n-1)}\sum_{i,j} \left(F_{ij} - O_i \frac{f\left(r_i,r_j,w_i,w_j,d_{ij}, s^{(r)}_{ij}, s^{(w)}_{ij}\right)}{\sum_{k\neq i}f\left(r_i,r_k,w_i,w_k,d_{ik}, s^{(r)}_{ik}, s^{(w)}_{ik}\right)}\right)^2 +\lambda C(f),
\label{eq:sr_objective}
\end{equation}
where $\mathcal{F}$ is the set of functions which can be formed by composition of the allowed operators; $n$ is the number of spatial units; $C(f)$ is the complexity of the function $f$;    $\lambda$ is a penalty factor for model complexity. Such a trade-off between model accuracy and simplicity is typical in statistical model selection \citep{BuAn02}. Furthermore, if we formulate SR as a multi-objective optimization problem regarding accuracy and simplicity, changing the parameter $\lambda$ would lead to different models on the Pareto frontier. 

SR is proven to be NP-hard \citep{ViPi22}, which indicates an exact and efficient solution probably does not exist. As a result, the heuristic genetic programming approach,  \citep{Koz94,ScLi09,LCOB21} remains the mainstream solution to SR, though methods based on physical principles \citep{UdTe20,UTF20}, neural networks \citep{SLM18,PLM21,KdAL22}, and Bayesian computation \citep{GRAM20,JFK20} are emerging recently. 

\par We develop our SR program based on a Julia package for genetic programming based SR, SymbolicRegression.jl \citep{Cran23}, as it is specially designed for scientific data analysis. Instead of penalizing the complexity of each expression (Eq.\ref{eq:sr_objective}), this software applies a penalty on complexity levels if they are too frequent in the population. In other words, it aims to balance explorations for models at a range of complexity (from 1 to 19 in this study). As a result, it can output the whole Pareto frontier in a single run. Moreover, expression simplification and constant optimization are included in the set of mutation operations to improve the models' compactness and accuracy. Several other enhancements to the general genetic programming framework are implemented, including the integration of simulated annealing, age-regularization to avoid premature convergence, and migration of expressions between several parallel populations. We refer to \citet{Cran23} for more technical details. 

Nevertheless, optimization of the allocation weight function is not readily supported by the package. We modified the evaluation module in SymbolicRegression.jl to calculate flow volumes from allocation weights. As constant optimization only occurs with probability, we also added a constant optimization step before the program exits, ensuring all output formulas have optimized parameter values. 

\par Due to the randomness of genetic programming, we repeat the SR procedure multiple times for each dataset. Typically, we first use all features as explanatory variables, then restrict the set of variables to those shown to be important. Fewer variables lead to a smaller search space, thus may yield better solutions. The number of iterations is set as 100 throughout this study. Batching is necessary to accelerate the program for large flow datasets. By default, the Broyden-Fletcher-Goldfarb-Shanno (BFGS) algorithm is used to optimize constants, and the Nelder-Mead algorithm is used if the former fails. The Pareto optimal expressions within the combined output of repeated runs are selected as final models. The detail of SR settings on each dataset is given in Supplementary Table~\ref{tab:srsetting}. 

\section{Results}
\subsection{Discovering symbolic models of mobility flows}

\begin{figure}[tp!]
\centering
\includegraphics[width=\linewidth]{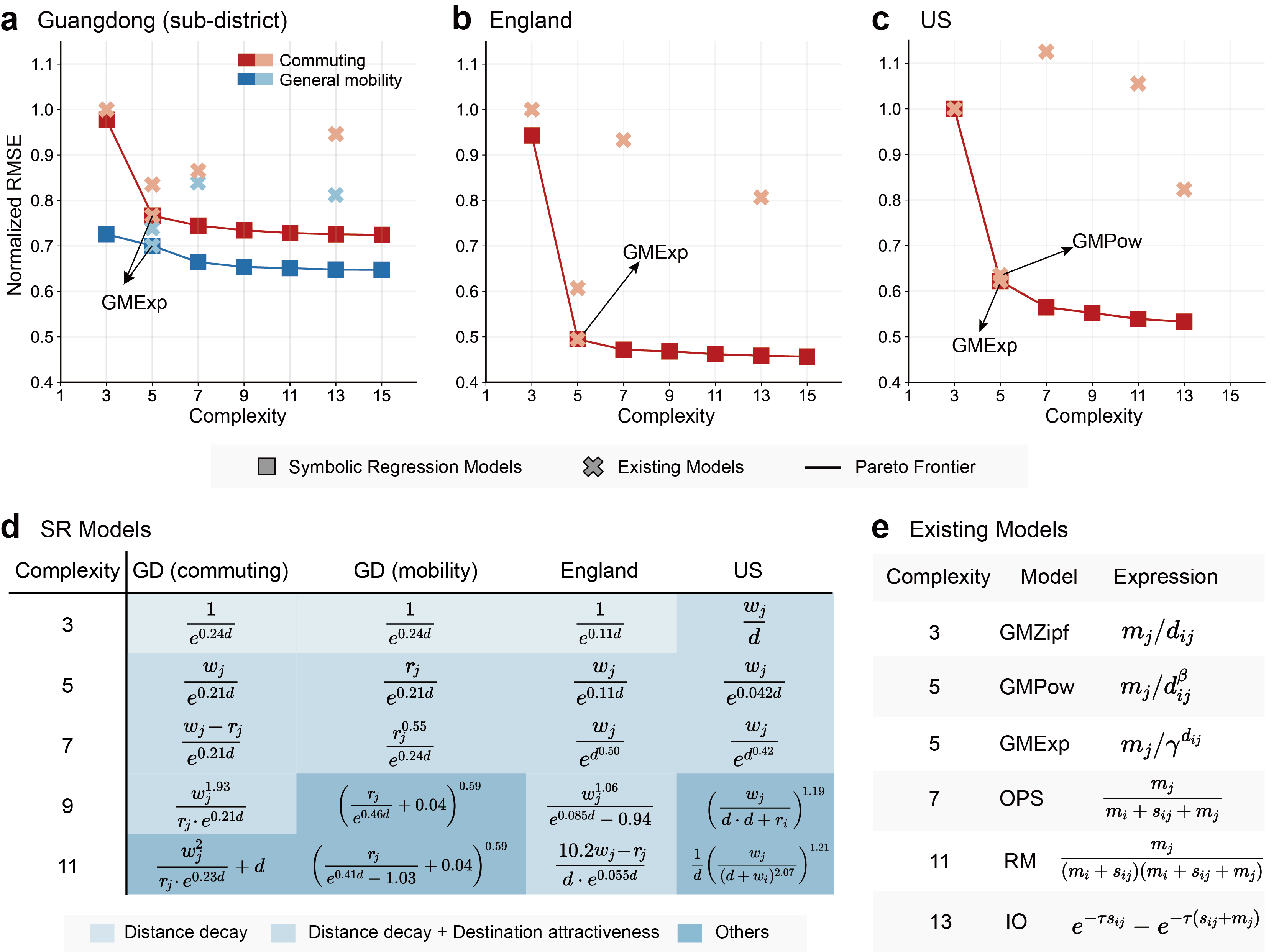}
\caption{\fontsize{11}{15}\linespread{1}\selectfont{\textbf{SR results on mobility flow data.}} (\textbf{a-c}) Pareto frontiers of SR models on Guangdong, England, and US datasets. As flow magnitudes vary across datasets, we normalize the RMSE with that of the simplest gravity model ($w_j/d_{ij}$). The accuracy and complexity of six existing models are marked with crosses (note that some existing models are not shown as their errors exceed the range of the y-axis). Expressions with complexity 1 are excluded as one variable or constant is not sufficient to model the mobility flow. The expressions with complexity 15 on the US dataset is not Pareto optimal. Across all datasets and complexity levels, the SR models consistently outperform or match the existing models in accuracy. (\textbf{d}) Expressions of the Pareto optimal SR models. The notations follow Fig.~\ref{fig:framework}, except that $d$ is short for $d_{ij}$. To align with existing models, some expressions are not given in the form with the lowest complexity. The distilled models are classified based on the captured effects governing human mobility. (\textbf{e}) Expressions of the considered existing models. 
}
\label{fig:srmain}
\end{figure}

As shown in Fig.~\ref{fig:srmain} and Supplementary Fig.~\ref{fig:srextend}, SR models achieve superior or equal accuracy to existing models of the same complexity, with existing models consistently falling on the upper-right side of the Pareto frontiers formed by the SR models (Fig.~\ref{fig:srmain} a-c). Notably, at a complexity of 5, SR autonomously discovered the gravity model with exponential decay in six of the seven datasets. In the remaining case, the accuracy of the exponential decay gravity model was almost identical to that of the model on the Pareto frontier (only 1.6\% higher RMSE; see Supplementary Table~\ref{tab:alter_gdcom}). This strong convergence across diverse datasets underscores the effectiveness and parsimony of the exponential decay gravity model for representing human mobility. Consistent with previous empirical analysis \citep{LBR16}, the distance decay parameters of the SR-derived gravity models show a decreasing trend with the area of the spatial units, from 0.21 for Guangdong sub-districts to 0.04 for US counties (Table \ref{tab:dataset}). Theoretically, Wilson's maximum entropy derivation of the exponential decay ($1/e^{c_{ij}}$, where $c_{ij}$ is the travel cost from location $i$ to $j$) assumes travel cost is proportional to distance (see Appendix). Our empirical findings provide strong support for this theoretical interpretation: by exploring a broad space of functional forms, SR identifies the exponential decay gravity model as a near-optimal solution balancing predictive power and model simplicity. The proximity of this model to the Pareto frontier’s inflection point further suggests that the exponential decay gravity model effectively captures the distribution of flow data.

Beyond the exponential decay, SR discovered a novel distance decay function, $f(d_{ij})=e^{d_{ij}^\alpha}$, at the complexity of 7 in both the England and US commuting datasets, with $\alpha = 0.50$ and $0.42$, respectively (Fig. \ref{fig:srmain}d). This new form substantially improves predictive accuracy, reducing RMSE by 4.6\% in the England data and 9.3\% in the US data. Within Wilson's maximum entropy framework mentioned above, this decay form corresponds to a power-law relationship between travel cost and distance ($c_{ij} \propto d_{ij}^\alpha$, see Appendix). While a linear relationship is commonly assumed in transportation research, the relationship between travel cost (including monetary and time cost) and distance can be complicated by mixed transportation modes, leading to different function forms \citep{YHW13}. Our finding provides a new hypothesis for empirical validation in the future. Regardless, the improved predictive performance of this interpretable decay function highlights its potential for future applications in human mobility modeling.

Regarding the other existing models, the gravity model with power-law decay is second best overall, with an average 11.2\% higher RMSE than the exponential decay (Table~\ref{tab:baseline_mse}). Models based on intervening opportunity do not perform well on most datasets. On average, the RMSE of the radiation model is 92.0\% higher than the SR model with the same complexity, and the error increase for Schneider's intervening opportunity model and opportunity priority selection model are 39.8\% and 49.1\%, respectively. Besides, the intervening opportunity appears in the Pareto optimal SR models on only one of the seven datasets. Hence, we found no evidence that the intervening opportunity better describes the decay effect than geographic distance.

\begin{table}
\centering
\caption{\fontsize{11}{15}\linespread{1}\selectfont{\textbf{The accuracy of existing models on flow data, measured by root mean squared error (RMSE).}} The lowest error on each dataset is put in bold. See Fig.~\ref{fig:srmain}e for model definitions. BTH is short for Beijing-Tianjin-Hebei urban agglomeration, China.}
\renewcommand{\arraystretch}{1.25} 
\small
\begin{tabular}{c|c|c|c|c|c|c|c}
\hline
Type & \multicolumn{4}{c|}{Commuting} & \multicolumn{3}{c}{Mobility}\\
\hline
Region & \multicolumn{2}{c|}{Guangdong} & England & US & \multicolumn{2}{c|}{Guangdong} & BTH\\
\hline
Granularity & sub-district & county & MLAD & county & sub-district & county & county\\
\hline
GMZipf & 15.02 & 146.25 & 728.58 & 2105.9 & 1588.5 & 12870 &81813\\
GMPow & 12.53 & 123.78 & 442.23 & 1337.7& 1173.6 & 11734 &76959\\
GMExp & \textbf{11.51} & 111.91 & \textbf{360.34} & \textbf{1309.9}& \textbf{1112.8} & \textbf{10723} &\textbf{64746}\\
RM & 22.72 & \textbf{105.62} & 863.67 & 2222.4 & 2023.1 & 14628 &119256\\
IO & 14.20 & 108.49 & 587.86 & 1733.4& 1289.8 & 12260 &74732\\
OPS & 12.99 & 131.69 & 679.65 & 2370.0& 1332.9 &13582 &85542\\
\hline
\end{tabular} 
\label{tab:baseline_mse}
\end{table}

\subsection{Geographical heterogeneity of mobility models}

\begin{figure}[p!]
\centering
\includegraphics[width=\linewidth]{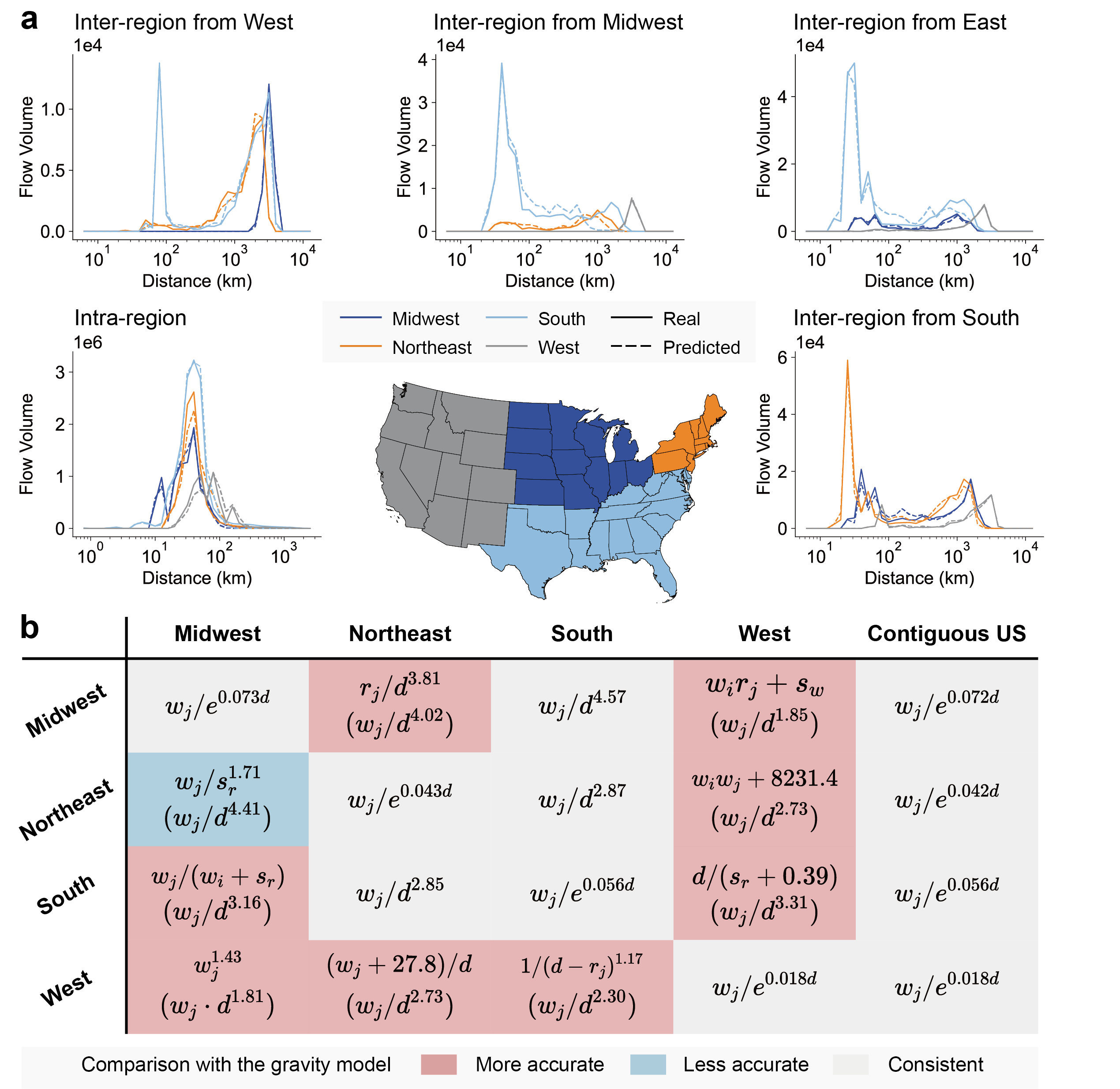}
\caption{\fontsize{11}{15}\linespread{1}\selectfont{\textbf{Spatial heterogeneity of the mobility model across US.}} (\textbf{a}) The distance distribution of commuting flows, grouped by geographic regions. The predicted flows are from the complexity 5 SR model on each subset grouped by the origin and destination region. For inter-region flows, each subplot shows outflows from a region, and the line color corresponds to the destination region. (\textbf{b}) SR models at complexity 5 for commuting flows (we report the complexity 3 model for flows from West to Midwest as the complexity 5 model is not Pareto optimal). Each row/column represents the region containing the residential/work place. The notations in formulas are the same as in Fig.~\ref{fig:framework}, except that $d$ is short for $d_{ij}$; $s_r, s_w$ are short for $s^{(r)}_{ij}, s^{(w)}_{ij}$, respectively. The accuracy of these models is compared with one of the power-law and exponential gravity models which produces lower error on the subset of flow data. A red/blue cell indicates the accuracy of the SR model is better/worse (measured with RMSE); in such cases, the gravity model is given in parentheses. A grey cell means the SR model is identical to the gravity model. }\label{fig:sphet}
\end{figure}

Given the nature of spatial heterogeneity, the global modeling approach discussed above may represent merely the average generation process of human mobility, ignoring possibly important spatial disparities \citep{Foth81,KHJ23,Yu24}. To investigate the consistency of mobility models across geographical areas, we partition the US commuting flow data according to the four regions (Northeast, Midwest, East, and West; Fig.~\ref{fig:sphet}a), resulting in 4 intra-region subsets and 12 inter-region subsets, and independently apply SR to each subset. Other SR settings are consistent with the analysis at the whole country level. 

For all four subsets of intra-region flows, the exponential decay gravity model remains optimal as suggested by SR (the diagonal expressions in Fig.~\ref{fig:sphet}b). This provides the basis for the model's overall optimality as intra-region flows dominate commuting flows, containing 97.5\% of flow volumes. Yet the exponential decay gravity model appears to be suboptimal for inter-region commuting. Over the 12 subsets of inter-region flows, the power-law decay gravity model consistently outperforms exponential decay. Moreover, even better models are discovered by SR in 8 of the 12 subsets. 

\par This disparity in optimal model form may indicate difference in the mechanisms governing inter-region and intra-region commuting. Actually, the distance distribution of intra-region flows follow a typical distance decay pattern, while uncommon bimodality is observed for inter-region subsets, with a long-range peak at around 1000~km (Fig.~\ref{fig:sphet}a). Through examination of the flow data, we find the destinations of these flows are mostly large cities, with the top 5 destinations as Chicago, New York, Los Angeles, Houston, and Dallas. We speculate that, with more job opportunities, the attractiveness of these large cities may be much stronger than the distance friction effect in long-distance commuting, leading to weak or even disappeared distance decay. This difference is also reflected in some SR models for inter-region commuting, where flow volumes do not decrease with distance as commonly assumed (e.g. models for flows between Midwest and West in Fig.~\ref{fig:sphet}b). 

\subsection{Stability of Symbolic Regression under Noise}

All observational data contain noise, and if the noise is high enough, it is likely that we will not be able to identify valid information or models \citep{FFRdLR23}. To evaluate the effectiveness of SR on noisy mobility data, we conducted experiments on synthetic data (since the generation model and noise structure in the real data are unknown). We align the settings with our analysis on US data. Using the real spatial boundary, workplace population, and total outflow, we generate the expected values of origin-destination flows $\mu_{ij}$ based on allocation weights determined by three mobility models: the exponential decay gravity model, the power-law decay gravity model, and radiation model (Eq.\ref{eq:normalize}, Fig.\ref{fig:srmain}e). We then apply a negative binomial model to represent random fluctuations, or factors not captured by the symbolic models \citep{Yu24}. Although the Poisson distribution is more commonly used in flow modeling \citep{SiTe06}, its variance is determined by the mean, while the negative binomial distribution with two parameters allows us to adjust the amount of random noise. Specifically, the probability mass function is given by
\begin{equation}
P(T_{ij}=k) = \frac{\Gamma(k+\rho_{ij})}{k!\Gamma(\rho_{ij})} \left(\frac{\mu_{ij}}{\rho_{ij}+\mu_{ij}}\right)^k \left(\frac{\rho_{ij}}{\rho_{ij}+\mu_{ij}}\right)^{\rho_{ij}},
\label{eq:negbin}
\end{equation}
where $\mu_{ij}, \rho_{ij}$ are parameters. It follows that 
\begin{equation}
    \mathrm{E}(T_{ij})=\mu_{ij},  \mathrm{Var}(T_{ij})=\mu_{ij}+\frac{\mu_{ij}^2}{\rho_{ij}}.
\end{equation}
We denote $\alpha_{ij}=\frac{1}{\rho_{ij}}$, and use the same value $\alpha_{ij} = \alpha$ for all origin-destination pairs. Hence, $\mathrm{Var}(T_{ij})=\mu_{ij}+\alpha \mu_{ij}^2$, so higher $\alpha$ indicates larger random noise. We consider a series of $\alpha$ values from 0 to 2.0 with a step of 0.25 (we use $\alpha = 10^{-8}$ as the minimal level as $\alpha=0$ is not allowed). For each $\alpha$ value, we sample each flow $T_{ij}$ independently from the negative binomial model. Finally, we rounded $T_{ij}$ to the closest integer. We generated five samples of flow data for each combination of 9 $\alpha$ values and three mobility models. The SR program was independently applied to each sample (see Table S1 for details), where we dropped flows less than 3 due to computational burden from a large number of such small flows. 

To illustrate the amount of noise in the synthetic data, we apply CPC to measure the discrepancy between flows without noise $\{\mu_{ij}\}$ and flows with noise $\{T_{ij}\}$. We use the CPC value of the exponential decay gravity model (the best-fit baseline model) on the real US commuting flows, 0.728, as a standard, reflecting the potential amount of random fluctuation in the real data. For all three considered models, the synthetic data reach a comparable CPC at $\alpha=0.5$ (Supplementary Table \ref{tab:compare_cpc}). 

\begin{figure}[ht!]
\centering
\includegraphics[width=8cm]{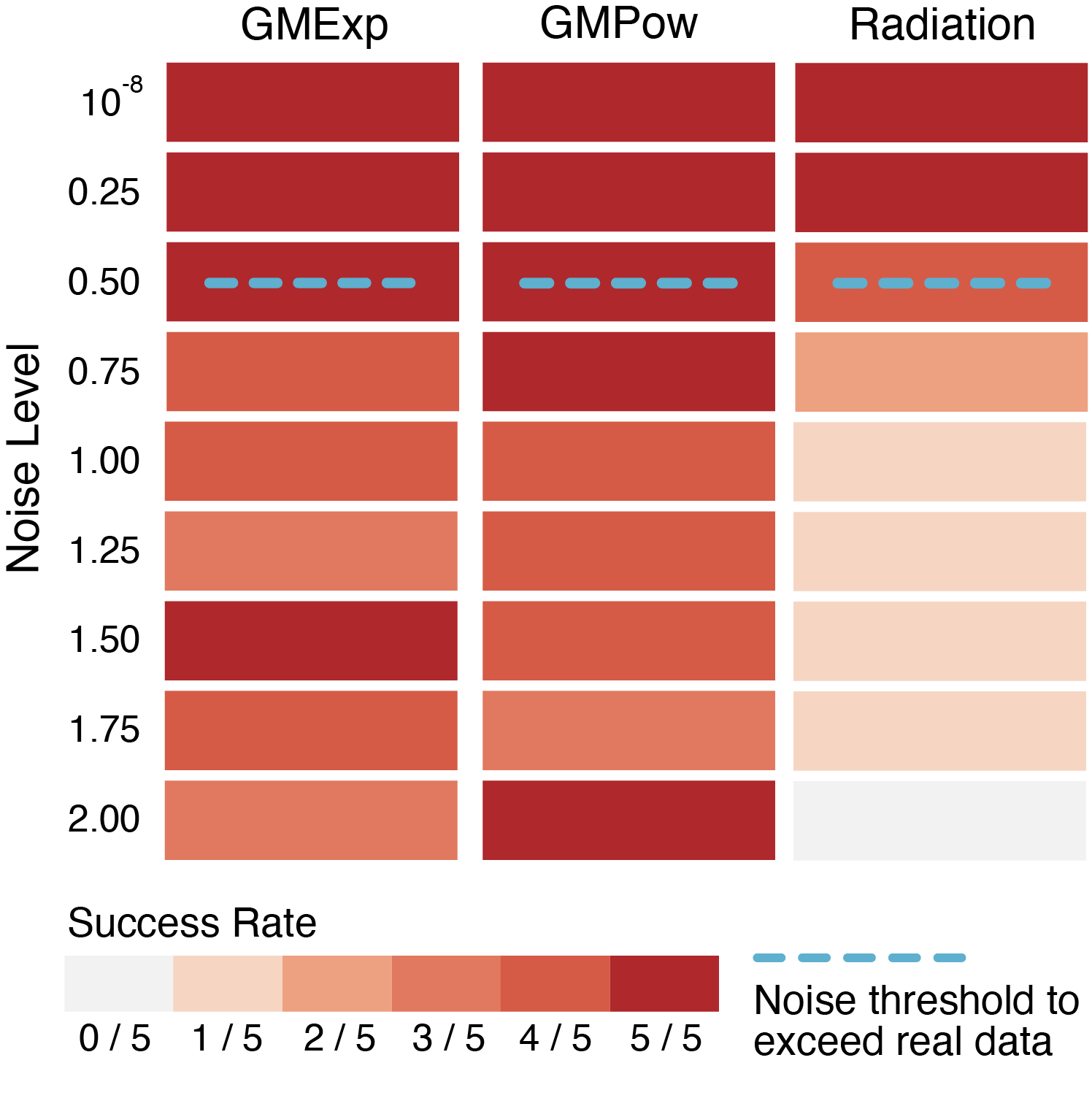}
\caption{\fontsize{11}{15}\linespread{1}\selectfont{\textbf{The success rate of SR to identify generation models on simulated data.}} The negative binomial distribution is applied to model the random fluctuations in mobility flows. The noise level comparable to that in real data (measured by model CPC) is marked with dashed lines. }\label{fig:simulated}
\end{figure}

Figure~\ref{fig:simulated} reports the number of successful instances where SR discovers the correct form of the generation model. When little random noise is applied, all three models can be successfully distilled. Furthermore, under a moderate amount of noise (more than $\sigma=0.5$), SR still discovers the two gravity models in most cases, while the discovery of the radiation model becomes unstable as the noise level increases. The latter, however, does not imply that SR produces sub-optimal models. In all but one among the 25 cases where the SR model with complexity 11 is inconsistent with the radiation model, the SR models actually have lower RMSE. This indicates the model that best describes the data has changed because of the noise introduced. Although such ``shifted'' formulas discovered by SR are inconsistent across samples, some of them resembles the radiation model, such as $w_j/((w_i+s_w)^{\beta}+w_j^\gamma)$, $w_j/((w_i+s_w)^\beta(w_i+w_j))$, $w_j/(w_i+s_w+w_j/w_i)^\beta$, where $\beta,\gamma$ are constants. These formulas suggest potential ways to extend the radiation model to better model noisy data. 

\section{Discussion and Conclusions}

Genetic programming-based SR not only enables us to examine the optimality of existing models, but also sheds light on how key factors are progressively incorporated as the allowed model complexity increases. As shown in Fig.~\ref{fig:srmain}d and Supplementary Fig.~\ref{fig:srextend}c, SR first captures the distance friction effect——near destinations attract more people than distant ones——at complexity 3, suggesting it is the primary effect governing human mobility. Another key factor revealed by SR is the destination attractiveness (represented by population), which is mostly incorporated into the models at complexity 5. Places with larger populations tend to attract more mobility flows, as they provide a greater number and diversity of goods, services, and job opportunities. Most SR models at complexity 7 and 9 share the form of $g(w_j,r_j)/f(d_{ij})$, which still combine the destination attractiveness and the distance friction effect, yet with more complex function forms for one or both effects. As complexity further increases, the SR models become inconsistent across datasets and difficult to explain. These models are not likely to be general models for mobility flows, but capturing the characteristics of a single dataset. 

Recently, \citet{CTDM25} applied Bayesian Machine Scientist (BMS) \citep{GRAM20}, a Bayesian-based approach to Symbolic Regression, to analyze human mobility data in the US. Their study demonstrated that models derived from BMS are at least comparable to complex machine learning models in predicting mobility flows, with function forms resembling the gravity model. For comparison, we applied the BMS approach to the US and England commuting data, and report the resulting formulas in Supplementary Table S9. BMS tends to generate models of complexity over 20, while our genetic programming-based approach identifies optimized models across a range of complexity from 3 to 15, including those comparable to existing mobility models. In this sense, the two approaches can complement each other in exploring the accuracy-complexity Pareto frontier of symbolic models. Furthermore, by examining the complexity level at which each variable is included, our approach can identify the primary and secondary effects in the process under investigation, which is not feasible with BMS. 

Our analysis also reveals the limitation of genetic programming-based SR. Although the performance of SR on simple formulas (complexity$<$9) is generally satisfactory, the derived formulas for longer expressions are inconsistent across repeated runs, indicating that sub-optimal solutions are found. Existing literature also points out that SR is only robustly applicable to datasets with fewer than 10 variables \citep{WTVN23}, due to the exponentially growing search space. Incorporating neural networks and symbolic regression might be a feasible approach to discover more complicated models. For example, SR could be applied to separable modules of a neural network, and the distilled formulas are then combined into a complete expression \citep{CSGB20,SDC23,LLS24}. Future studies could investigate the effectiveness of such neuro-symbolic approaches for model building in human mobility.

In conclusion, this work demonstrates that SR is capable of discovering accurate and compact models that capture the regularities in human mobility, and the derived Pareto optimal models reveal feasible ways to extend the basic gravity model, some of which are previously unknown. In the future, it is intriguing to apply SR to more location features other than population for mobility model discovery. More generally, the accumulation of observational data and the progress in symbolic regression provide a promising way to build empirical models in the social sciences. Rather than comparing the accuracy of existing candidate models, SR allows us to answer the questions: Does an existing model have the best form among all possible expressions, given a set of operators? If not, does SR suggest new candidate models of theoretical interest? We look forward to applications of SR beyond mobility data, which may provide new perspectives on the quantitative modeling of social phenomena. 

\subsection*{Acknowledgments}
We thank Tianyou Cheng and Chengling Tang for assistance in processing the cellphone datasets; Andre Python, Ryan Q. Wang, Ce Hou, and Xuechen Wang for helpful discussion and comments.

\subsection*{Funding}
We acknowledge the support of the National Natural Science Foundation of China under Grant Nos. 42422110 and 42430106. L.D. was supported by the Fundamental Research Funds for the Central Universities, Peking University.

\subsection*{Declaration of competing interest}
The authors declare no competing interests.

\subsection*{Author contributions}
L.D., Y.L., and H.G. conceived the project and designed the experiments; H.G., W.Z., and Y.H. collected the data; H.G., W.Z., and J.Y. performed the experiments; W.Z., and H.G., visualized the results; H.G., L.D., W.Z, and Y.L. wrote the manuscript.

\subsection*{Data and code availability}
All codes and processed flow datasets from the US, the UK, and Beijing-Tianjin-Hebei, China are publicly available on GitHub at https://github.com/urbansci/FlowSR. The Guangdong cellphone dataset can be requested from the corresponding author.

\renewcommand{\theequation}{A.\arabic{equation}}
\setcounter{equation}{0}

\begin{appendices}

\section{Maximum entropy approach to the gravity model}

Inspired by statistical mechanics,   \citet{Wil70} applied the concept of entropy on origin-destination flow data, and deduced the gravity model using the maximum entropy principle. Let $F_{ij}$ denote the flow from place $i$ to $j$, the entropy of a flow distribution $\{F_{ij}\}$ is defined as
\begin{equation}
    S = -\sum_{i,j} p_{ij} \ln p_{ij}
\end{equation}
where $p_{ij}=F_{ij}/\sum_{i,j} F_{ij}$. Here the entropy is approximately linear to the logarithmic number of ``microscopic states'' of the system (a microscopic state contains a full description of the origin and destination of each person), hence the distribution $\{F_{ij}\}$ with maximum entropy corresponds to the most probable ``macroscopic state''. 
\par Consider the following constraints on mobility flows: 
\begin{align}
    \sum_j F_{ij}&=O_i, \forall i\\
    \sum_i F_{ij}&=D_j, \forall j\\
    \sum_{i,j} F_{ij}c_{ij}&=C
\end{align}
where $c_{ij}$ denote the travel cost from place $i$ to $j$. The first two constraints ensure the distribution is consistent with the total outflow and inflow of each place. The last constraint asserts that human mobility is restricted by travel costs. Using the method of Lagrange multipliers to maximize $S$ given the constraints above, we get the doubly-constrained gravity model as follows:
\begin{equation}
    F_{ij}=A_iB_jO_iD_j e^{-\beta c_{ij}}
\end{equation}
where $A_i = \left(\sum_j B_jD_j e^{-\beta c_{ij}}\right)^{-1}$, $B_j = \left(\sum_i A_iO_ie^{-\beta c_{ij}}\right)^{-1}$. This form suggests that flow volumes follow an exponential decay with travel cost. The decay parameter $\beta$ arises from the unknown total cost $C$, and usually needs real data to be calibrated. 

\par The decay function of distance can be deduced by substituting the relationship between travel cost and distance. For example, given $c_{ij}=kd_{ij}$, we get the exponential decay $e^{-k\beta d_{ij}}$; given $c_{ij}=k\ln d_{ij}$, we get the power-law decay $d_{ij}^{-k\beta}$. Therefore, the discovered exponential-power decay $e^{-d_{ij}^\alpha}$ indicates a power-law relationship between travel cost and distance $c_{ij}=\frac{1}{\beta}d_{ij}^\alpha$.

\end{appendices}

\printbibliography

@article{And11,
author = {Anderson, James E.},
title = {The gravity model},
journal = {Annual Review of Economics},
volume = {3},
number = {1},
pages = {133-160},
year = {2011}
}

@article{AHS24,
title={The 15-minute city quantified using human mobility data},
author={Abbiasov, Timur and Heine, Cate and Sabouri, Sadegh and Salazar-Miranda, Arianna and Santi, Paolo and Glaeser, Edward and Ratti, Carlo},
journal={Nature Human Behaviour},
volume={8},
number={3},
pages={445--455},
year={2024}
}

@article{BBG18,
title = {Human mobility: Models and applications},
journal = {Physics Reports},
volume = {734},
pages = {1-74},
year = {2018},
author = {Hugo Barbosa and Marc Barthelemy and Gourab Ghoshal and Charlotte R. James and Maxime Lenormand and Thomas Louail and Ronaldo Menezes and José J. Ramasco and Filippo Simini and Marcello Tomasini}
}

@article{BdLNE15,
author = {Barbosa, Hugo and de Lima-Neto, Fernando B. and Evsukoff, Alexandre and Menezes, Ronaldo},
title = {The effect of recency to human mobility},
journal = {EPJ Data Science},
volume = {4},
number = {1},
pages = {21},
year = {2015}
}

@ARTICLE{CBB20,
AUTHOR={Cardoso, Pedro and Branco, Vasco V. and Borges, Paulo A. V. and Carvalho, José C. and Rigal, François and Gabriel, Rosalina and Mammola, Stefano and Cascalho, José and Correia, Luís},   
TITLE={Automated discovery of relationships, models, and principles in ecology}, 
JOURNAL={Frontiers in Ecology and Evolution}, 
VOLUME={8}, 
YEAR={2020}
}

@article{CLG16,
author = {Çolak, Serdar and Lima, Antonio and González, Marta C.},
title = {Understanding congested travel in urban areas},
journal = {Nature Communications},
volume = {7},
number = {1},
pages = {10793},
year = {2016}
}

@article{CTDM25,
author = {Cabanas-Tirapu, Oriol and Danús, Lluís and Moro, Esteban and Sales-Pardo, Marta and Guimerà, Roger},
title = {Human mobility is well described by closed-form gravity-like models learned automatically from data},
journal = {Nature Communications},
volume = {16},
number = {1},
pages = {1336},
year = {2025}
}

@article{Dip98,
author = {G Diplock},
title ={Building New Spatial Interaction Models by Using Genetic Programming and a Supercomputer},
journal = {Environment and Planning A: Economy and Space},
volume = {30},
number = {10},
pages = {1893-1904},
year = {1998}
}

@article{FFRdLR23,
author = {Fajardo-Fontiveros, Oscar and Reichardt, Ignasi and De Los Ríos, Harry R. and Duch, Jordi and Sales-Pardo, Marta and Guimerà, Roger},
title = {Fundamental limits to learning closed-form mathematical models from data},
journal = {Nature Communications},
volume = {14},
number = {1},
pages = {1043},
year = {2023}
}

@article{Foth81,
author = {A. Stewart Fotheringham},
title ={Spatial structure and distance-decay parameters},
journal = {Annals of the Association of American Geographers},
volume = {71},
number = {3},
pages = {425--436},
year = {1981}
}

@article{GRAM20,
author = {Roger Guimerà  and Ignasi Reichardt  and Antoni Aguilar-Mogas  and Francesco A. Massucci  and Manuel Miranda  and Jordi Pallarès  and Marta Sales-Pardo },
title = {A {Bayesian} machine scientist to aid in the solution of challenging scientific problems},
journal = {Science Advances},
volume = {6},
number = {5},
pages = {eaav6971},
year = {2020},
}

@article{JLY20,
author = {Jia, Jayson S. and Lu, Xin and Yuan, Yun and Xu, Ge and Jia, Jianmin and Christakis, Nicholas A.},
title = {Population flow drives spatio-temporal distribution of {COVID-19} in {China}},
journal = {Nature},
volume = {582},
number = {7812},
pages = {389-394},
year = {2020}
}

@article{KHJ23,
author = {Kwon, Oh-Hyun and Hong, Inho and Jung, Woo-Sung and Jo, Hang-Hyun},
title = {Multiple gravity laws for human mobility within cities},
journal = {EPJ Data Science},
volume = {12},
number = {1},
pages = {57},
year = {2023}
}

@article{Koz94,
author = {Koza, John R.},
title = {Genetic programming as a means for programming computers by natural selection},
journal = {Statistics and Computing},
volume = {4},
number = {2},
pages = {87-112},
year = {1994}
}

@article{LBR16,
title = {Systematic comparison of trip distribution laws and models},
journal = {Journal of Transport Geography},
volume = {51},
pages = {158-169},
year = {2016},
author = {Maxime Lenormand and Aleix Bassolas and José J. Ramasco}
}

@article{LHF25,
title = {Generating equitable urban human flows with a fairness-aware deep learning model},
journal = {Cities},
volume = {167},
pages = {106296},
year = {2025},
author = {Zhewei Liu and Lipai Huang and Chao Fan and Ali Mostafavi},
}

@article{LHG12,
author = {Lenormand, Maxime and  Huet, Sylvie and Gargiulo, Floriana and Deffuant, Guillaume},
journal = {PLOS ONE},
title = {A universal model of commuting networks},
year = {2012},
month = {10},
volume = {7},
pages = {1-7},
number = {10},
}

@article{LiYa19,
title = {New parameter-free mobility model: Opportunity priority selection model},
journal = {Physica A: Statistical Mechanics and its Applications},
volume = {526},
pages = {121023},
year = {2019},
author = {Erjian Liu and Xiaoyong Yan}
}

@article{LJC23,
year = {2023},
volume = {4},
number = {4},
pages = {045002},
author = {Pablo Lemos and Niall Jeffrey and Miles Cranmer and Shirley Ho and Peter Battaglia},
title = {Rediscovering orbital mechanics with machine learning},
journal = {Machine Learning: Science and Technology}
}

@article{LLS24,
author = {Liu, Sannyuya and Li, Qing and Shen, Xiaoxuan and Sun, Jianwen and Yang, Zongkai},
title = {Automated discovery of symbolic laws governing skill acquisition from naturally occurring data},
journal = {Nature Computational Science},
volume = {4},
number = {5},
pages = {334-345},
year = {2024}
}

@article{Lill89,
author = {Eduard Lill},
title = {Die Grundgesetze des Personenverkehrs},
journal = {Zeitschrift für Eisenbahnen und Dampfschiffahrt der Österreichisch-Ungarischen Monarchie},
volume = {35},
pages = {697--706},
year = {1889},
}

@article{LZW24,
title = {Generating sparse origin–destination flows on shared mobility networks using probabilistic graph neural networks},
journal = {Sustainable Cities and Society},
volume = {114},
pages = {105777},
year = {2024},
author = {Yuebing Liang and Zhan Zhao and Chris Webster}
}

@article{MaCa24,
author = {Makke, Nour and Chawla, Sanjay},
title = {Interpretable scientific discovery with symbolic regression: A review},
journal = {Artificial Intelligence Review},
volume = {57},
number = {1},
pages = {2},
year = {2024}
}

@article{PMS23,
author = {Pappalardo, Luca and Manley, Ed and Sekara, Vedran and Alessandretti, Laura},
title = {Future directions in human mobility science},
journal = {Nature Computational Science},
volume = {3},
number = {7},
pages = {588-600},
year = {2023}
}

@article{Open88,
author = {Openshaw, Stan},
title = {Building an Automated Modeling System to Explore a Universe of Spatial Interaction Models},
journal = {Geographical Analysis},
volume = {20},
number = {1},
pages = {31-46},
year = {1988}
}

@article{Rave85,
author = {E. G. Ravenstein},
journal = {Journal of the Statistical Society of London},
number = {2},
pages = {167--235},
title = {The Laws of Migration},
volume = {48},
year = {1885}
}

@article{RoTh04,
title = {Spatial interaction modelling},
journal = {Papers in Regional Science},
volume = {83},
number = {1},
pages = {339-361},
year = {2004},
author = {John R. Roy and Jean Claude Thill}
}

@article{RPSP20,
title = {Bayesian machine scientist to compare data collapses for the Nikuradse dataset},
author = {Reichardt, Ignasi and Pallar\`es, Jordi and Sales-Pardo, Marta and Guimer\`a, Roger},
journal = {Physical Review Letters},
volume = {124},
issue = {8},
pages = {084503},
numpages = {5},
year = {2020},
publisher = {American Physical Society}
}

@article{SBB23,
title={{COVID-19} is linked to changes in the time--space dimension of human mobility},
author={Santana, Clodomir and Botta, Federico and Barbosa, Hugo and Privitera, Filippo and Menezes, Ronaldo and Di Clemente, Riccardo},
journal={Nature Human Behaviour},
volume={7},
number={10},
pages={1729--1739},
year={2023}
}

@Article{SBL21,
author = {Filippo Simini and Gianni Barlacchi and Massimilano Luca and Luca Pappalardo},
title = {A Deep Gravity model for mobility flows generation},
journal = {Nature Communications},
year = {2021},
volume = {12},
pages = {6576}
}

@article{SDOK21,
  title={The universal visitation law of human mobility},
  author={Schl{\"a}pfer, Markus and Dong, Lei and O’Keeffe, Kevin and Santi, Paolo and Szell, Michael and Salat, Hadrien and Anklesaria, Samuel and Vazifeh, Mohammad and Ratti, Carlo and West, Geoffrey B},
  journal={Nature},
  volume={593},
  number={7860},
  pages={522--527},
  year={2021},
  publisher={Nature Publishing Group UK London}
}

@article{Sch59,
author = {Schneider, Morton},
title = {Gravity models and trip distribution theory},
journal = {Papers in Regional Science},
volume = {5},
number = {1},
pages = {51-56},
year = {1959}
}

@article{ScLi09,
author = {Michael Schmidt and Hod Lipson},
title = {Distilling free-form natural laws from experimental data},
journal = {Science},
volume = {324},
number = {5923},
pages = {81-85},
year = {2009}
}

@article{SGM12,
author = {Simini, Filippo and González, Marta C. and Maritan, Amos and Barabási, Albert-László},
title = {A universal model for mobility and migration patterns},
journal = {Nature},
volume = {484},
number = {7392},
pages = {96-100},
year = {2012},
}

@article{SiTe06,
    author = {Silva, J. M. C. Santos and Tenreyro, Silvana},
    title = {The Log of Gravity},
    journal = {The Review of Economics and Statistics},
    volume = {88},
    number = {4},
    pages = {641-658},
    year = {2006}
}

@article{SKW10,
author = {Song, Chaoming and Koren, Tal and Wang, Pu and Barabási, Albert-László},
title = {Modelling the scaling properties of human mobility},
journal = {Nature Physics},
volume = {6},
number = {10},
pages = {818-823},
year = {2010}
}

@article{Stew41,
author = {John Q. Stewart},
journal = {Science},
number = {2404},
pages = {89--90},
title = {An Inverse Distance Variation for Certain Social Influences},
volume = {93},
year = {1941}
}

@article{Stou40,
author = {Samuel A. Stouffer},
journal = {American Sociological Review},
number = {6},
pages = {845--867},
title = {Intervening opportunities: A theory relating mobility and distance},
volume = {5},
year = {1940}
}

@article{UdTe20,
author = {Silviu-Marian Udrescu  and Max Tegmark },
title = {{AI Feynman}: A physics-inspired method for symbolic regression},
journal = {Science Advances},
volume = {6},
number = {16},
pages = {eaay2631},
year = {2020}
}

@article{ViPi22,
title={Symbolic Regression is {NP}-hard},
author={Marco Virgolin and Solon P Pissis},
journal={Transactions on Machine Learning Research},
year={2022},
url={https://openreview.net/forum?id=LTiaPxqe2e}
}

@article{VNAAG21,
year = {2021},
volume = {915},
number = {1},
pages = {71},
author = {Francisco Villaescusa-Navarro and Daniel Anglés-Alcázar and Shy Genel and David N. Spergel and Rachel S. Somerville and Romeel Dave and Annalisa Pillepich and Lars Hernquist and Dylan Nelson and Paul Torrey and Desika Narayanan and Yin Li and Oliver Philcox and Valentina La Torre and Ana Maria Delgado and Shirley Ho and Sultan Hassan and Blakesley Burkhart and Digvijay Wadekar and Nicholas Battaglia and Gabriella Contardo and Greg L. Bryan},
title = {The {CAMELS} Project: cosmology and astrophysics with machine-learning simulations},
journal = {The Astrophysical Journal}
}

@article{WKS19,
author = {Wang, Jinzhong and Kong, Xiangjie and Xia, Feng and Sun, Lijun},
title = {Urban Human Mobility: Data-Driven Modeling and Prediction},
year = {2019},
volume = {21},
number = {1},
journal = {ACM SIGKDD Explorations Newsletter},
pages = {1–19},
numpages = {19}
}

@article{WTVN23,
author = {Digvijay Wadekar  and Leander Thiele and Francisco Villaescusa-Navarro  and J. Colin Hill  and Miles Cranmer  and David N. Spergel  and Nicholas Battaglia  and Daniel Anglés-Alcázar  and Lars Hernquist  and Shirley Ho },
title = {Augmenting astrophysical scaling relations with machine learning: Application to reducing the {Sunyaev–Zeldovich} flux–mass scatter},
journal = {Proceedings of the National Academy of Sciences},
volume = {120},
number = {12},
pages = {e2202074120},
year = {2023}
}

@article{WSW21,
title = {Simulating the urban spatial structure with spatial interaction: A case study of urban polycentricity under different scenarios},
journal = {Computers, Environment and Urban Systems},
volume = {89},
pages = {101677},
year = {2021},
author = {Cai Wu and Duncan Smith and Mingshu Wang}
}

@article{WSZ20,
author = {Weng, Baicheng and Song, Zhilong and Zhu, Rilong and Yan, Qingyu and Sun, Qingde and Grice, Corey G. and Yan, Yanfa and Yin, Wan-Jian},
title = {Simple descriptor derived from symbolic regression accelerating the discovery of new perovskite catalysts},
journal = {Nature Communications},
volume = {11},
number = {1},
pages = {3513},
year = {2020}
}

@ARTICLE{YGZ21,
author={Yao, Xin and Gao, Yong and Zhu, Di and Manley, Ed and Wang, Jiaoe and Liu, Yu},
journal={IEEE Transactions on Intelligent Transportation Systems}, 
title={Spatial Origin-Destination Flow Imputation Using Graph Convolutional Networks}, 
year={2021},
volume={22},
number={12},
pages={7474-7484}
}

@article{YHW13,
author = {Yan, Xiao-Yong and Han, Xiao-Pu and Wang, Bing-Hong and Zhou, Tao},
title = {Diversity of individual mobility patterns and emergence of aggregated scaling laws},
journal = {Scientific Reports},
volume = {3},
pages = {2678},
year = {2013}
}

@article{Yu24,
author = {Hanchen Yu},
title = {Exploring multiscale spatial interactions: Multiscale geographically weighted negative binomial regression},
journal = {Annals of the American Association of Geographers},
volume = {114},
number = {3},
pages = {574--590},
year = {2024}
}

@article{Zipf46,
author = {George Kingsley Zipf},
journal = {American Sociological Review},
number = {6},
pages = {677--686},
title = {The P1 P2/D hypothesis: On the intercity movement of persons},
volume = {11},
year = {1946}
}

@book{BuAn02,
author  = {Kenneth P. Burnham and David R. Anderson},
publisher = {Springer},
title = {Model Selection and Multimodel Inference: A Practical Information-Theoretic Approach},
year = {2002},
edition = {Second Edition}
}

@book{Wil70,
author = {A. G. Wilson},
publisher  = {Routledge},
title = {Entropy in Urban and Regional Modelling},
year = {1970},
address = {London}
}

@inproceedings{CSGB20,
author = {Cranmer, Miles and Sanchez-Gonzalez, Alvaro and Battaglia, Peter and Xu, Rui and Cranmer, Kyle and Spergel, David and Ho, Shirley},
editor = {H. Larochelle and M. Ranzato and R. Hadsell and M.F. Balcan and H. Lin},
booktitle = {Advances in Neural Information Processing Systems},
pages = {17429--17442},
title = {Discovering Symbolic Models from Deep Learning with Inductive Biases},
volume = {33},
year = {2020}
}

@inproceedings{KdAL22,
author = {Kamienny, Pierre-Alexandre and d'Ascoli, St\'{e}phane and Lample, Guillaume and Charton, Francois},
booktitle = {Advances in Neural Information Processing Systems},
editor = {S. Koyejo and S. Mohamed and A. Agarwal and D. Belgrave and K. Cho and A. Oh},
pages = {10269--10281},
title = {End-to-end Symbolic Regression with Transformers},
volume = {35},
year = {2022}
}

@inproceedings{LCOB21,
author = {La Cava, William and Orzechowski, Patryk and Burlacu, Bogdan and de Franca, Fabricio and Virgolin, Marco and Jin, Ying and Kommenda, Michael and Moore, Jason},
booktitle = {Proceedings of the Neural Information Processing Systems Track on Datasets and Benchmarks},
editor = {J. Vanschoren and S. Yeung},
title = {Contemporary Symbolic Regression Methods and their Relative Performance},
volume = {1},
year = {2021}
}

@Inproceedings{LMX20,
author  = {Zhicheng Liu and Fabio Miranda and Weiting Xiong and Junyan Yang and Qiao Wang and Claudio T. Silva}, 
title   = {Learning Geo-Contextual Embeddings for Commuting Flow Prediction}, 
booktitle = {Proceedings of the AAAI Conference on Artificial Intelligence},
year = {2020},
volume = {34},
pages = {808-816}
}

@inproceedings{PLM21,
title={Deep symbolic regression: Recovering mathematical expressions from data via risk-seeking policy gradients},
author={Brenden K Petersen and Mikel Landajuela Larma and Terrell N. Mundhenk and Claudio Prata Santiago and Soo Kyung Kim and Joanne Taery Kim},
booktitle={Proceedings of the 9th International Conference on Learning Representations},
year={2021},
url={https://openreview.net/forum?id=m5Qsh0kBQG}
}

@inproceedings{SDC23,
title={Learning Symbolic Models for Graph-structured Physical Mechanism},
author={Hongzhi Shi and Jingtao Ding and Yufan Cao and Quanming Yao and Li Liu and Yong Li},
booktitle={Proceedings of the 11th International Conference on Learning Representations},
year={2023},
url={https://openreview.net/forum?id=f2wN4v\_2\_\_W}
}

@inproceedings{UTF20,
author = {Udrescu, Silviu-Marian and Tan, Andrew and Feng, Jiahai and Neto, Orisvaldo and Wu, Tailin and Tegmark, Max},
editor = {H. Larochelle and M. Ranzato and R. Hadsell and M.F. Balcan and H. Lin},
booktitle = {Advances in Neural Information Processing Systems},
pages = {4860--4871},
title = {{AI Feynman} 2.0: Pareto-optimal symbolic regression exploiting graph modularity},
volume = {33},
year = {2020}
}

@InProceedings{SLM18,
title = {Learning Equations for Extrapolation and Control},
author = {Sahoo, Subham and Lampert, Christoph and Martius, Georg},
booktitle = {Proceedings of the 35th International Conference on Machine Learning},
pages = {4442--4450},
year = 	 {2018},
editor = {Dy, Jennifer and Krause, Andreas},
volume = {80},
series = {Proceedings of Machine Learning Research}
}

@misc{Cran23, 
author		= "Cranmer, M.",
title			= "Interpretable Machine Learning for Science with PySR and SymbolicRegression.jl", 
year			= "2023",
note			= "Preprint at \url{https://arxiv.org/abs/2305.01582}"
}

@misc{JFK20,
title={Bayesian Symbolic Regression}, 
author={Ying Jin and Weilin Fu and Jian Kang and Jiadong Guo and Jian Guo},
year={2020},
note= "Preprint at \url{https://arxiv.org/abs/1910.08892}"
}

@misc{VeDo22,
title={Machine Learning the Gravity Equation for International Trade}, 
author={Sergiy Verstyuk and Michael R. Douglas},
year={2022},
note= "Preprint at \url{https://ssrn.com/abstract=4053795}"
}

\clearpage
\renewcommand{\thefigure}{S\arabic{figure}}
\renewcommand{\thetable}{S\arabic{table}}
\renewcommand{\theequation}{S\arabic{equation}}
\renewcommand{\thepage}{S\arabic{page}}
\setcounter{figure}{0}
\setcounter{table}{0}
\setcounter{equation}{0}
\setcounter{page}{1}

\begin{figure}[htbp]
\centering
\includegraphics{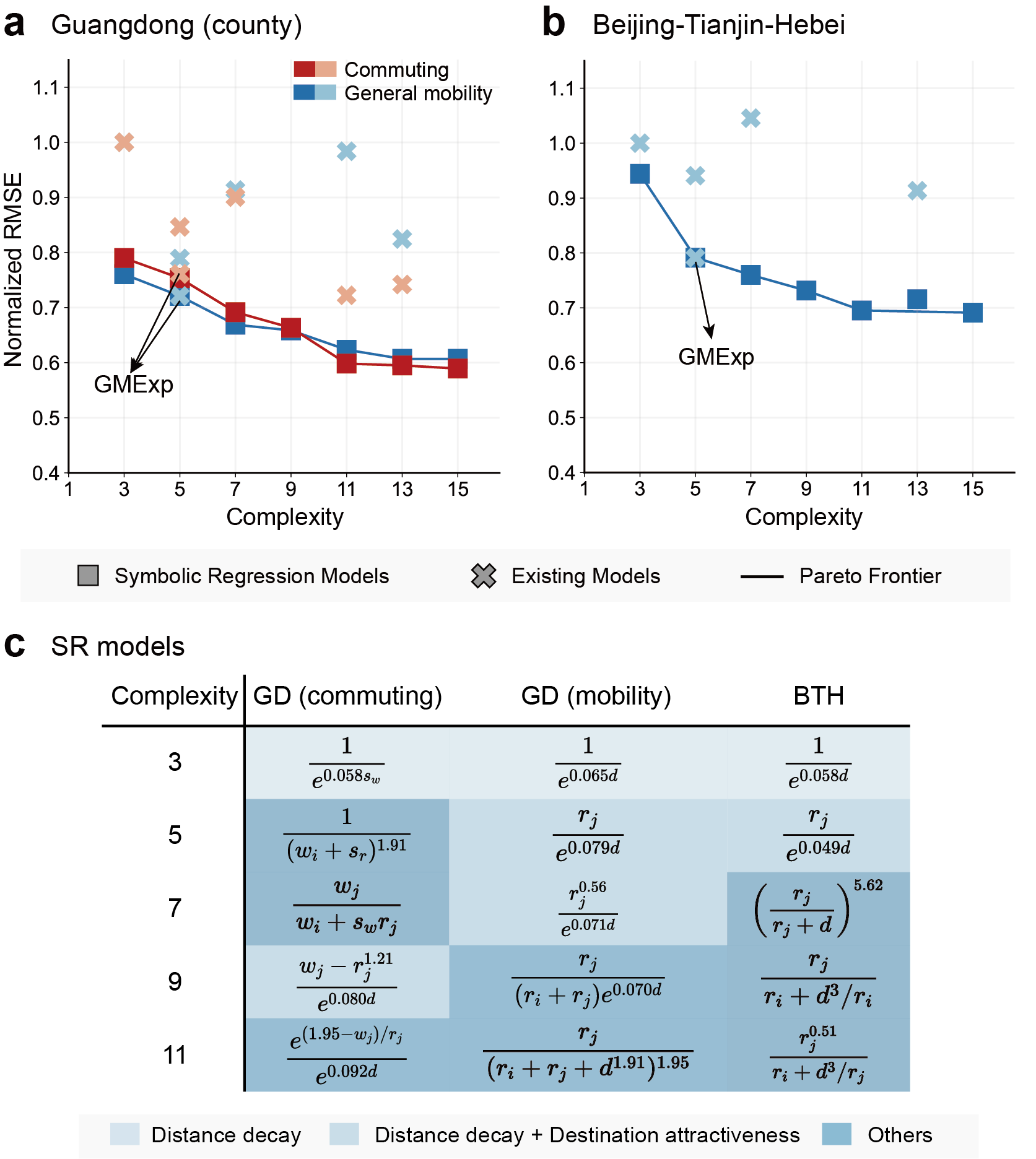}
\caption{\fontsize{11}{15}\linespread{1}\selectfont{\textbf{SR results on Guangdong (county level) and Beijing-Tianjin-Hebei (BTH) data. }}(\textbf{a-b}) Pareto frontiers of SR models on Guangdong and BTH data. As in Fig.~\ref{fig:srmain}, the RMSEs are normalized with that of the simplest gravity model ($w_j/d_{ij}$), and the crosses represent the accuracy and complexity of six existing models (Fig.~\ref{fig:srmain}e). (\textbf{c}) Expressions of the Pareto optimal SR models. The notations follow Fig.~\ref{fig:framework}, except that $d$ is short for $d_{ij}$. The distilled models are classified based on the captured effects governing human mobility. A decay function of intervening opportunity is viewed as a special form of distance decay. 
}
\label{fig:srextend}
\end{figure}
\clearpage

\begin{figure}
\centering
\includegraphics[width=\linewidth]{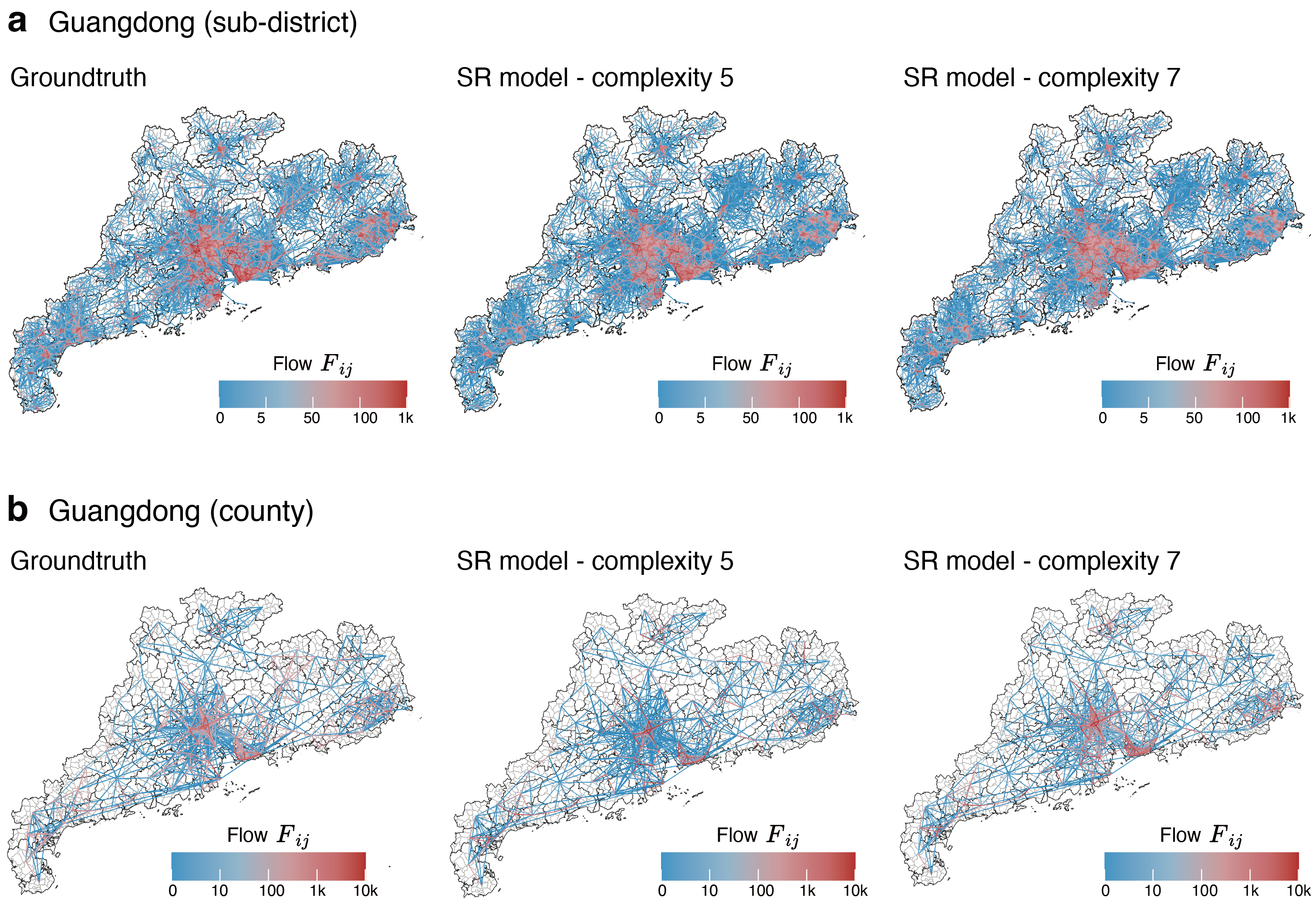}
\caption{\fontsize{11}{15}\linespread{1}\selectfont{\textbf{Actual and SR model-generated commuting flow maps in Guangdong, China, at the (\textbf{a}) sub-district and (\textbf{b}) county levels.}} SR models at complexity 5 and 7 effectively capture the overall mobility patterns, particularly in estimating flows with high volumes.}
\label{fig:flow_gdcom}
\end{figure}
\clearpage

\begin{figure}
\centering
\includegraphics[width=\linewidth]{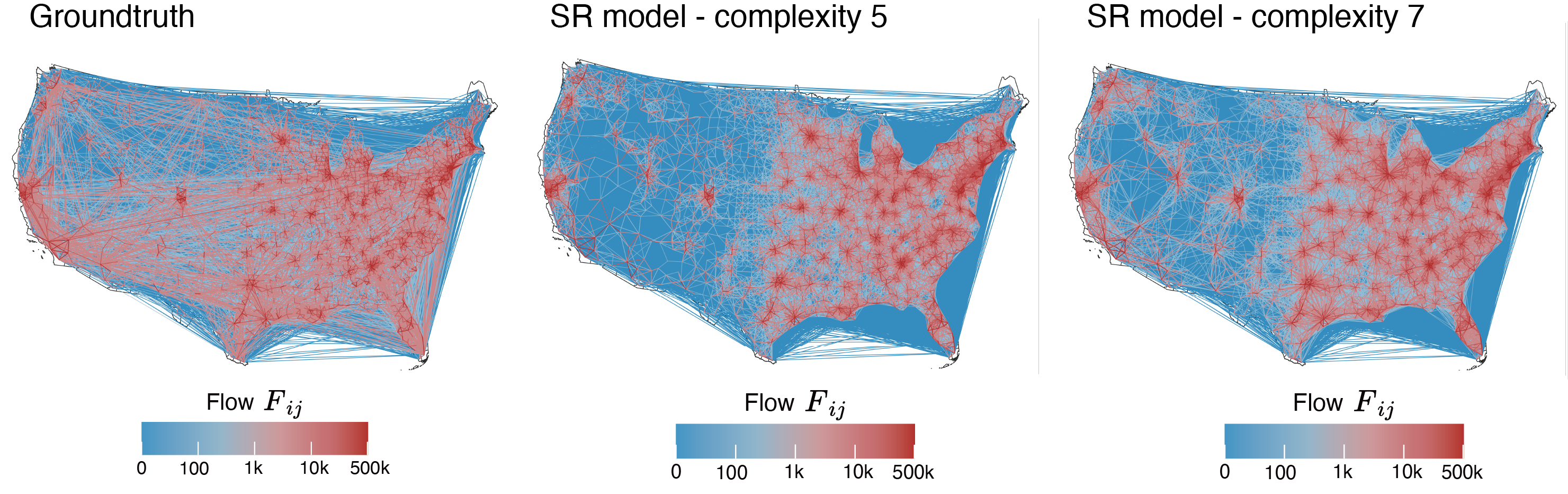} 
\caption{\fontsize{11}{15}\linespread{1}\selectfont{\textbf{Actual and SR model-generated commuting flow maps in the US. }}The SR models with complexity 5 and 7 are used.}
\label{fig:flow_us}
\end{figure}

\begin{figure}
\centering
\includegraphics[width=\linewidth]{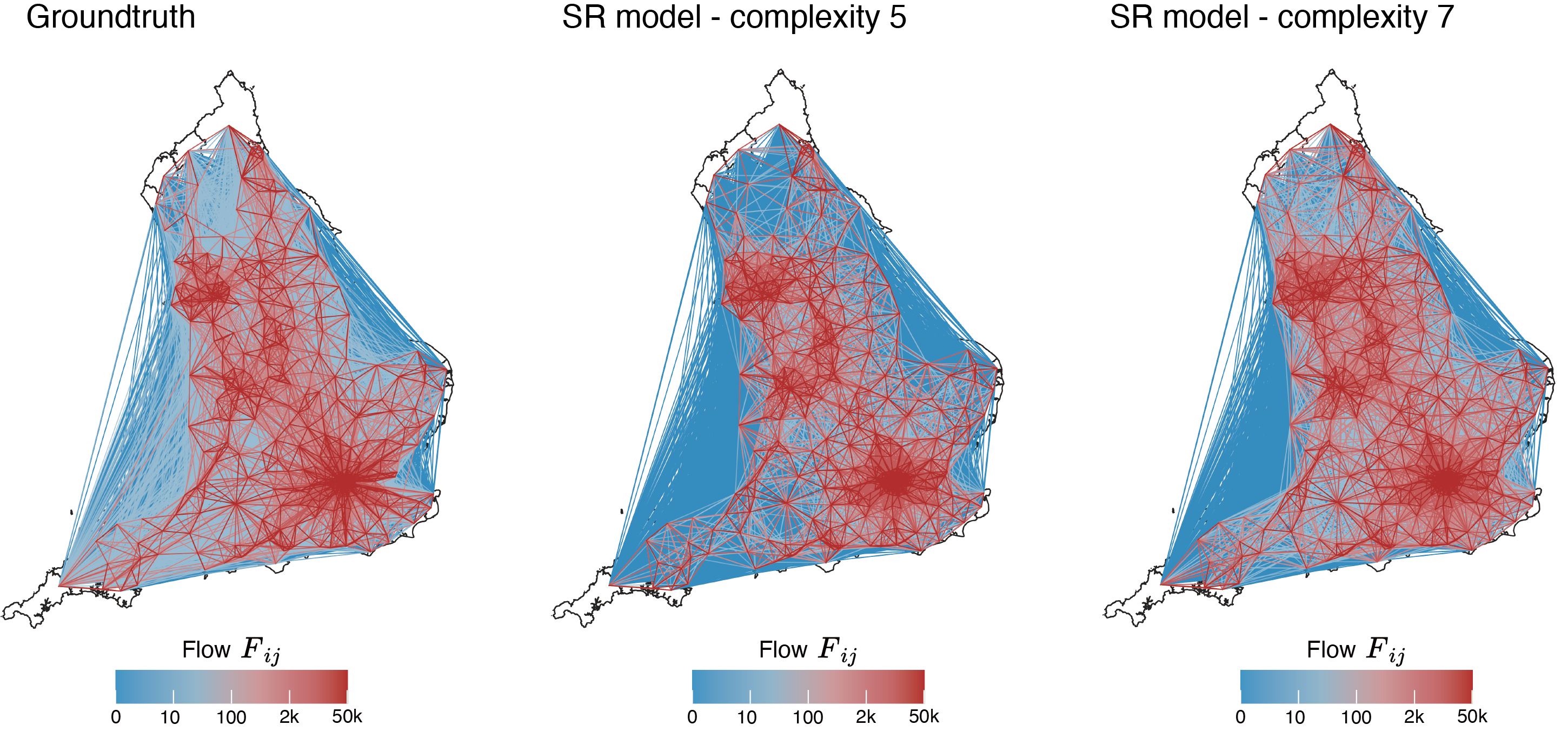} 
\caption{\fontsize{11}{15}\linespread{1}\selectfont{\textbf{Actual and SR model-generated commuting flow maps in England. }} The SR models with complexity 5 and 7 are used.}
\label{fig:flow_uk}
\end{figure}
\clearpage

\begin{figure}
\centering
\includegraphics[width=\linewidth]{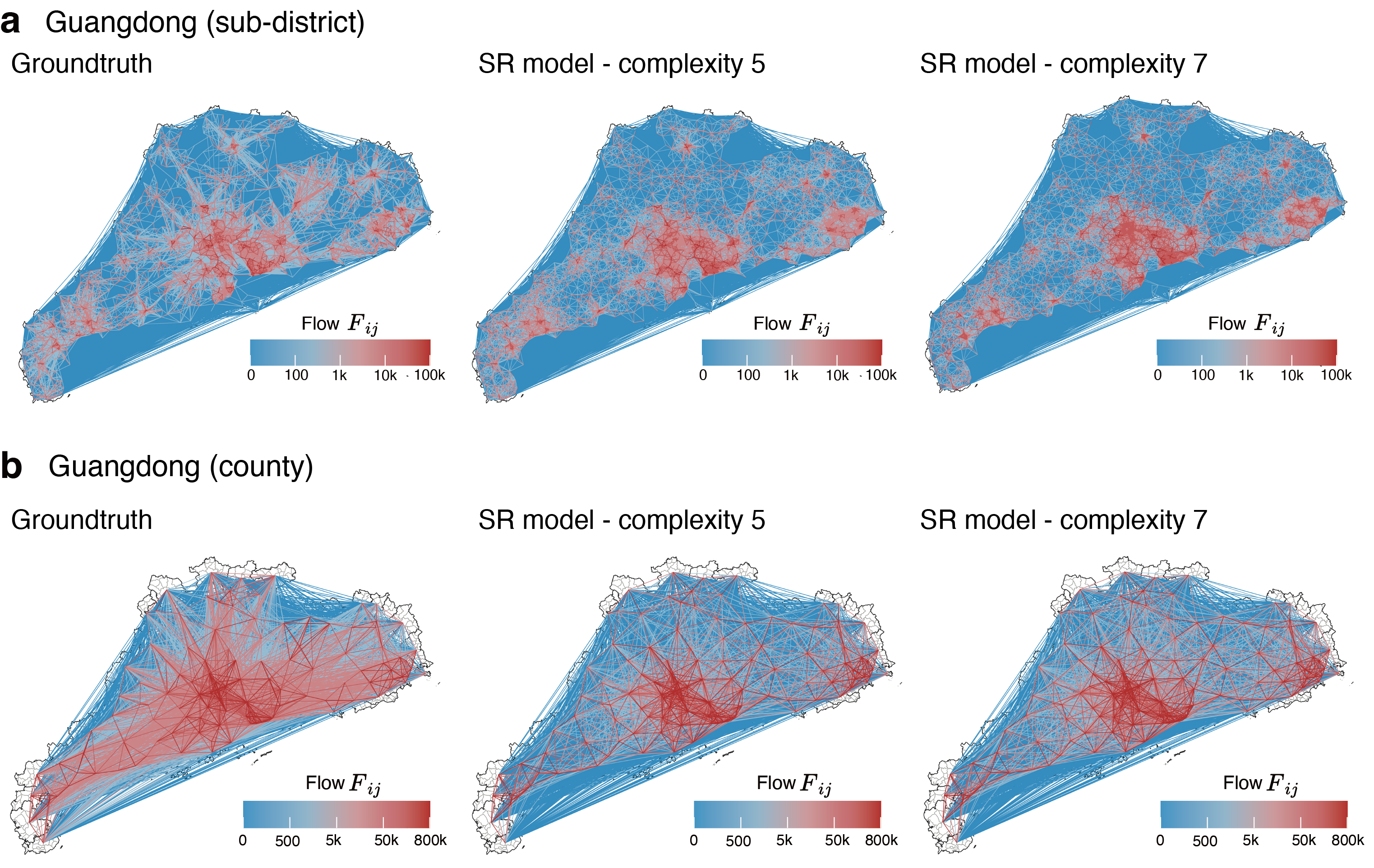}
\caption{\fontsize{11}{15}\linespread{1}\selectfont{\textbf{Actual and SR model-generated general mobility flow maps in Guangdong, China, at the (\textbf{a}) sub-district and (\textbf{b}) county scale. }} The SR models with complexity 5 and 7 are used.}
\label{fig:flow_gdmob}
\end{figure}

\begin{figure}
\centering\includegraphics[width=\linewidth]{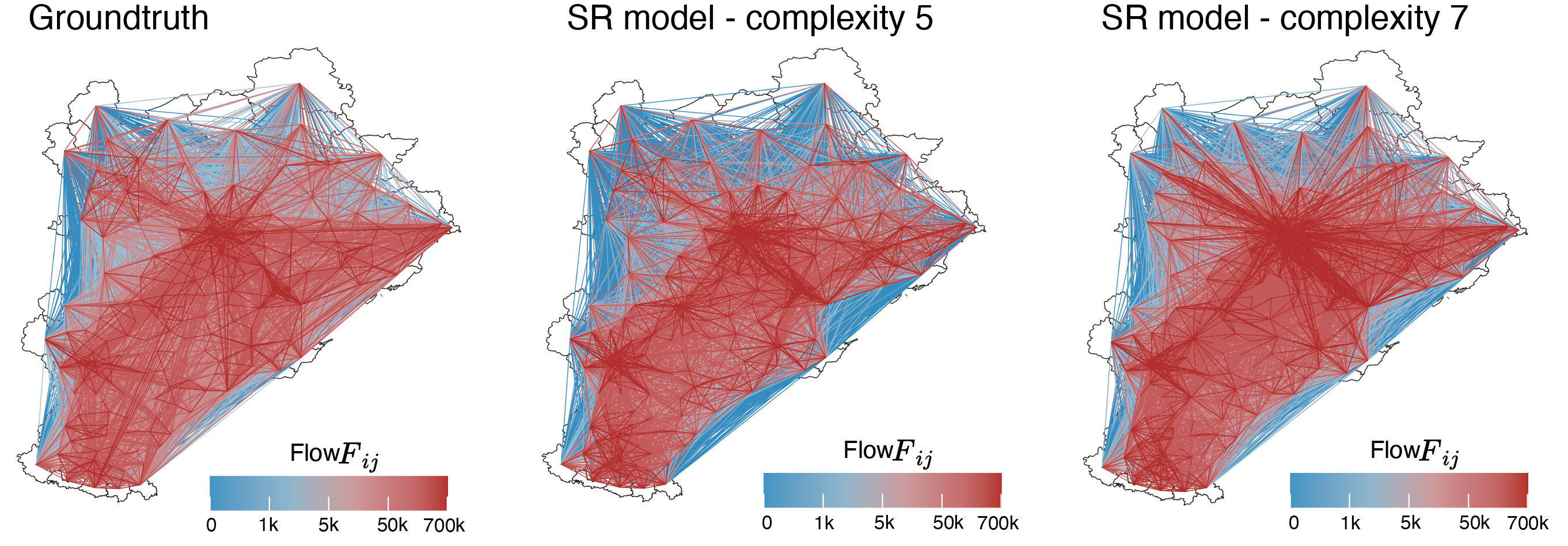}
\caption{\fontsize{11}{15}\linespread{1}\selectfont{\textbf{Actual and SR model-generated general mobility flow maps in Beijing-Tianjin-Hebei urban agglomeration. }} The SR models with complexity 5 and 7 are used.}
    \label{fig:flow_bth}
\end{figure}

\begin{table}
\centering
\caption{\fontsize{11}{15}\linespread{1}\selectfont{\textbf{Symbolic Regression settings.} opt means the algorithm for constant optimization. By default, batching is not used, and opt=    ``BFGS''. BTH is short for Beijing-Tianjin-Hebei urban agglomeration, China.}}
\renewcommand{\arraystretch}{1.0}
\begin{tabular}{cccccc}
\hline
Region & Granularity & Type &  Explanatory Variables & \#Runs & Other Parameters \\
\hline
\multirow{2}{*}{Guangdong} & \multirow{2}{*}{sub-district} & \multirow{2}{*}{commuting} & $r_i, r_j, w_i, w_j, d, s_r, s_w$ & 5 & \multirow{2}{*}{batchsize=40} \\
& & & $w_i, w_j, d$ & 5 & \\
\hline
\multirow{2}{*}{Guangdong} & \multirow{2}{*}{sub-district} & \multirow{2}{*}{mobility} & $r_i, r_j, d, s_r$ & 5 & \multirow{2}{*}{batchsize=40} \\
& & & $r_i, r_j, d$ & 5 & \\
\hline
\multirow{3}{*}{Guangdong}& \multirow{3}{*}{county} & \multirow{3}{*}{commuting} & $r_i, r_j, w_i, w_j, d, s_r, s_w$ & 5 & \\
& & & $w_i, w_j, r_j, d$ & 5 & \\
& & & $w_i, w_j, r_j, s_w$ & 5 & \\
\hline

\multirow{2}{*}{Guangdong} & \multirow{2}{*}{county} & \multirow{2}{*}{mobility } & $r_i, r_j, d, s_r$ & 5 & \\
& & & $r_i, r_j, d$ & 5 & \\
\hline
\multirow{2}{*}{England} & \multirow{2}{*}{MLAD} & \multirow{2}{*}{commuting} & $r_i, r_j, w_i, w_j, d, s_r, s_w$ & 5 & \\
& & & $w_i, w_j, d$ & 5 & \\
\hline
\multirow{2}{*}{US} & \multirow{2}{*}{county} & \multirow{2}{*}{commuting} & $r_i, r_j, w_i, w_j, d, s_r, s_w$ & 5 & \multirow{2}{*}{\shortstack{batchsize=100\\opt=``Nelder-Mead''}} \\
& & & $w_i, w_j, d$ & 5 & \\
\hline
US & county & synthetic & $w_i, w_j, d, s_w$ & 1 & \shortstack{batchsize=100\\opt=``Nelder-Mead''}\\
\hline

BTH & county & mobility & $r_i, r_j, d, s_r$ & 5 & \\
\hline
\end{tabular} 
\label{tab:srsetting}
\end{table}
\clearpage

\begin{table}
\centering
\caption{\fontsize{11}{15}\linespread{1}\selectfont{\textbf{Alternative SR models on Guangdong commuting data.} The top 3 distinct expressions on each complexity level are shown. If results across repeated runs are identical, there may be fewer than 3 expressions to show. }}
\renewcommand{\arraystretch}{1.0} 
\begin{tabular}{ccccc}
\hline
Dataset & Complexity & Expression & RMSE & CPC\\
\hline
\multirow{9}{*}{Guangdong (sub-district)} & 3 & $0.79^d$ & 14.68& 0.647\\
& 5 & $w_j/{1.24}^d$ & 11.51& 0.715\\
& 7 & $(w_j-r_j)/{1.24}^d$ & 11.18& 0.719\\
& 7 & $w_j/\exp({d^{0.56}})$ & 11.41& 0.722\\
& 9 & $w_j^{1.93}/(r_j\cdot{1.24}^d)$ & 11.03& 0.723\\
& 9 & $w_j\cdot w_j/(r_j\cdot{1.24}^d)$ & 11.06 & 0.719\\
& 11 & $w_j^2/(r_j\cdot{1.26}^d)+d$ & 10.94& 0.730\\
& 11 & $w_j^2/(r_j \exp({d^{0.57}}))$ & 11.01& 0.727\\
& 11 & $w_j(w_j+0.025) / (r_j\cdot{1.25}^d)$ & 11.02 & 0.719\\
\hline
\multirow{13}{*}{Guangdong (county)} & 3 &$0.94^{s_w}$ & 115.50 & 0.640\\
& 3 &$0.93^d$ & 123.85 & 0.641\\
& 5 & $(w_i+s_r)^{-1.91}$ & 110.18 & 0.660\\
& 5 & $(w_i+s_w)^{-1.32}$ & 110.76 & 0.659\\
& 5 & $w_j/1.10^d$ & 111.90 & 0.655\\
& 7 & $w_j/(w_i+s_wr_j)$ & 101.15 & 0.657\\
& 7 & $(w_j-r_j)/1.09^d$ & 104.97 & 0.670\\
& 9 & $(w_j-r_j^{1.21})/1.09^d$ & 97.00 & 0.716\\
& 9 & $((w_i+s_w)/w_j+r_j)^{-2.01}$ & 97.08 & 0.703\\
& 9 & $w_j / (1.34 ^ {r_j} \cdot s_w) + w_i)$ & 97.32 & 0.682\\
& 11 & $e^{(w_j-1.95)/r_j}/{1.10^{d}}$ & 87.49 & 0.727\\
& 11 & $w_j^2 \cdot (0.6 ^ {w_j})/ (1.09 ^ d)$ & 91.35& 0.648\\
& 11 & $((w_j+s_w)/(w_j-r_i)+ r_j)^{-1.80}$ & 91.77& 0.677\\
\hline
\end{tabular}
\label{tab:alter_gdcom} 
\end{table}
\clearpage

\begin{table}
\centering
\caption{\fontsize{11}{15}\linespread{1}\selectfont{\textbf{Alternative SR models on US commuting data.} The top 3 distinct expressions on each complexity level are shown. If results across repeated runs are identical, there may be fewer than 3 expressions to show. }}
\renewcommand{\arraystretch}{1.0} 
\begin{tabular}{ccccc}
\hline
Dataset & Complexity & Expression & RMSE & CPC\\
\hline
\multirow{12}{*}{US} & 3 &$w_j/d$ & 2105.9 & 0.407\\
& 5 & $w_j/{1.04}^{d}$ & 1309.9 & 0.728\\
& 5 & $w_j/d^{2.05}$ & 1337.7 & 0.664\\
& 7 & $w_j/\exp(d^{0.42})$ & 1188.6 & 0.724\\
& 7 & $w_j/(d\cdot{1.02}^d)$ & 1217.8 & 0.717\\
& 7 & $w_j/(d^{2.20}+w_i)$ & 1231.8 & 0.686\\
& 9 & $(w_j/(d\cdot d+r_i))^{1.19}$ & 1154.3 & 0.690\\
& 9 & $w_j / (d ^ {2.86} + 67.7w_i)$ & 1165.0 & 0.743\\
& 9 & $w_j/(d+w_i/d)^{2.54}$ & 1167.7 & 0.721\\
& 11 & $(w_j / (d + w_i)^{2.07})^{1.21} / d$ & 1135.0 & 0.743\\
& 11 & $w_j /\exp((d + r_j / w_j)^{0.43})$ & 1144.3 & 0.728\\
& 11 & $(w_j / (d\cdot(d + w_i) ^ {1.93})) ^ {1.20}$ & 1146.0 & 0.746\\
\hline
\end{tabular}
\label{tab:alter_us} 
\end{table}

\begin{table}
\centering
\caption{\fontsize{11}{15}\linespread{1}\selectfont{\textbf{Alternative SR models on England commuting data.} The top 3 distinct expressions on each complexity level are shown. If results across repeated runs are identical, there may be fewer than 3 expressions to show. }}
\renewcommand{\arraystretch}{1.0} 
\begin{tabular}{ccccc}
\hline
Dataset & Complexity & Expression & RMSE & CPC\\
\hline
\multirow{11}{*}{England} & 3 & $0.89^d$ & 687.10 & 0.661 \\
& 5 & $w_j/{1.11}^d$ & 360.34 & 0.770\\
& 7 & $w_j/\exp(d^{0.50})$ & 343.62 & 0.784\\
& 7 & $w_j/({1.09}^d-0.92)$ & 344.62 & 0.779 \\
& 7 & $w_j/(d\cdot {1.06}^d)$ & 345.75 & 0.781\\
& 9 & $w_j^{1.06}/({1.09}^d-0.94)$ & 341.02 & 0.780 \\
& 9 & $(w_j / ({1.09} ^ d - 0.89)) ^ {1.04}$ & 341.41 & 0.780 \\
& 9 & $((w_j / {1.05} ^ d) ^ {1.07}) / d$ & 341.48 & 0.782 \\
& 11 & $(10.2 w_j  - r_j) / (d\cdot {1.06} ^ d)$ & 336.49 & 0.786 \\
& 11 & $w_j/((1.06 ^ d)(d + r_j / w_j))$ & 337.51 & 0.783 \\
& 11 & $(16.8w_j - r_j) / (1.09 ^ d -0.95)$ & 337.54 & 0.782 \\
\hline
\end{tabular}
\label{tab:alter_uk} 
\end{table}
\clearpage

\begin{table}
\centering
\caption{\fontsize{11}{15}\linespread{1}\selectfont{\textbf{Alternative SR models on Guangdong general mobility data.} The top 3 distinct expressions on each complexity level are shown. If results across repeated runs are identical, there may be fewer than 3 expressions to show. }}
\renewcommand{\arraystretch}{1.0}
\begin{tabular}{ccccc}
\hline
Dataset & Complexity & Expression & RMSE & CPC \\
\hline
\multirow{10}{*}{Guangdong (sub-district)} & 3 & $0.78^d$ & 1153.2 & 0.606\\
& 5 & $r_j/{1.24}^d$ & 1112.8 & 0.623\\
& 7 & $r_j^{0.55}/{1.27}^d$ & 1055.2& 0.633\\
& 7 & $(r_j/{1.51}^d)^{0.57}$ & 1055.4& 0.633\\
& 9 & $(r_j/{1.58}^d + 0.04)^{0.59}$ & 1038.6& 0.626\\
& 9 & $r_j ^ {0.54}/1.29 ^ d +0.07$ & 1039.3 & 0.633\\
& 9 & $(r_j/1.49 ^ d)^{0.67} +0.4$ & 1039.6& 0.626\\
& 11 & $(r_j/(1.51^d-1.03)+0.04)^{0.59}$ & 1034.4& 0.637\\
& 11 & $(r_j ^ {0.54}/1.29 ^ d+0.13)^{1.05}$ & 1038.3& 0.627\\
& 11 & $(0.53 \cdot r_j / 1.61^d)^{0.55}+0.06$ & 1038.5 & 0.630\\
\hline
\multirow{6}{*}{Guangdong (county)} & 3 &$0.93^d$ & 11308.9 & 0.667\\
& 5 & $r_j/1.08^{d}$ & 10723.1 & 0.681\\
& 7 & $r_j^{0.56}/{1.07}^d$ & 9948.1 & 0.694\\
& 9 & $(r_j/(r_j+r_i))/1.07^d$ & 9793.8& 0.697\\
& 11 & $r_j/(d^{1.91}+r_j+r_i)^{1.95} $ & 9271.2 & 0.717\\
& 11 & $r_j / (r_i + r_j+ d^2) ^ {1.73}$ & 9327.8& 0.713\\
\hline
\end{tabular}
\label{tab:alter_gdmob} 
\end{table}

\begin{table}
\centering
\caption{\fontsize{11}{15}\linespread{1}\selectfont{\textbf{Alternative SR models on Beijing-Tianjin-Hebei general mobility data.} The top 3 distinct expressions on each complexity level are shown. If results across repeated runs are identical, there may be less than 3 expressions to show. }}
\renewcommand{\arraystretch}{1.0}
\begin{tabular}{ccccc}
\hline
Dataset & Complexity & Expression & RMSE & CPC\\
\hline
\multirow{11}{*}{Beijing-Tianjin-Hebei } & 3 & $0.94^d$ & 77235 & 0.655 \\
& 5 & $r_j/{1.05}^d$ & 64746 & 0.682 \\
& 7 & $(r_j/(d+r_j))^{5.62}$ & 62173 & 0.670 \\
& 7 & $r_j / (d ^ {1.91} + r_i)$ & 63173 & 0.669 \\
& 7 & $(r_j + 18.98) / {1.05}^d$ & 64033 & 0.688 \\
& 9 & $r_j / (r_i + d^3/r_i)$ & 59010 & 0.720 \\
& 9 & $r_j / \exp(r_i ^ {-0.63} d)$ & 59843 & 0.704 \\
& 9 & $r_j / \exp((d + r_i) ^ {0.53})$ & 59883 & 0.712 \\
& 11 & $r_j ^ {0.51} / (d^3/r_j + r_i)$ & 56870 & 0.731 \\
& 11 & $r_j / (d / (r_i / d) ^ {1.75} +1.14)$ & 58777 & 0.720 \\
& 11 & $(r_j / \exp(r_i ^ {-0.60}d)) ^ {0.90}$ & 59643 & 0.704 \\
\hline
\end{tabular}
\label{tab:alter_bth} 
\end{table}
\clearpage

\begin{table}
\centering
\caption{\fontsize{11}{15}\linespread{1}\selectfont{\textbf{The accuracy of existing models on flow data, measured by common part of commuters (CPC).}} The best value on each dataset is put in bold. See Fig.~\ref{fig:srmain}e for model definitions. BTH is short for Beijing-Tianjin-Hebei urban agglomeration, China.}
\renewcommand{\arraystretch}{1.25} 
\small
\begin{tabular}{c|c|c|c|c|c|c|c}
\hline
Type & \multicolumn{4}{c|}{Commuting} & \multicolumn{3}{c}{Mobility}\\
\hline
Region & \multicolumn{2}{c|}{Guangdong} & England & US & \multicolumn{2}{c|}{Guangdong} & BTH\\
\hline
Granularity & sub-district & county & MLAD & county & sub-district & county & county\\
\hline
GMZipf &0.634 &0.540 & 0.461 & 0.407 &0.343 &0.425 & 0.483\\
GMPow &0.705 &0.635 & 0.723 & 0.664 &0.599 & 0.613& 0.558\\
GMExp &\textbf{0.715} &0.687 &\textbf{0.770} & \textbf{0.728} &\textbf{0.622} & \textbf{0.683}& \textbf{0.682}\\
RM &0.577 &\textbf{0.694} & 0.649 & 0.605 & 0.538& 0.653 & 0.591\\
IO &0.637 &0.663 & 0.628 & 0.526 & 0.523& 0.610& 0.594\\
OPS &0.691 &0.577 & 0.576 & 0.483 & 0.491& 0.489 & 0.501\\
\hline
\end{tabular} 
\label{tab:baseline_cpc}
\end{table}

\begin{table}
\centering
\caption{\fontsize{11}{15}\linespread{1}\selectfont{\textbf{The discrepancy between synthetic flows with/without noise, measured by common part of commuters (CPC).}} Average CPC over five random samples are reported. The CPC value which is closest to the real data (0.728) for each model is put in bold. See Fig.~\ref{fig:srmain}e for model definitions. $\alpha$ is the dispersion parameter in the negative binomial distribution.}
\renewcommand{\arraystretch}{1.25} 
\small
\begin{tabular}{c|ccccccccc}
\hline
$\alpha$& 1e-8 & 0.25 & 0.5 & 0.75 & 1.0 & 1.25 & 1.5 & 1.75 & 2.0\\
\hline
GMPow & 0.972	&	0.802	&	\textbf{0.723}	&	0.677	&	0.635	&	0.60	&	0.571	&	0.551	&	0.539\\
GMExp & 0.989	&	0.800	&	\textbf{0.730}	&	0.672	&	0.635	&	0.602	&	0.566	&	0.543	&	0.523\\
RM & 0.989	&	0.804	&	\textbf{0.725}	&	0.675	&	0.633	&	0.601	&	0.573	&	0.541	&	0.523\\
\hline
\end{tabular} 
\label{tab:compare_cpc}
\end{table}
\clearpage

\begin{table}
\centering
\caption{\fontsize{11}{15}\linespread{1}\selectfont{\textbf{SR models from Bayesian Machine Scientist (BMS) on US and England commuting data.}} The explanatory variables include population $w_i, w_j$ and distance $d$, and the target variable is $\ln F_{ij}$. We run five independent Markov chains of 12,000 Monte Carlo steps. For each run, a random sample of 1,000 origin-destination pairs is used as training data. Up to six constants $c_1,\dots,c_6$ are allowed. These settings are aligned with \citet{CTDM25}. BMS adopts description length as the objective function. The model with the minimum description length in each run is reported. }
\renewcommand{\arraystretch}{1.25} 
\small
\begin{tabular}{c|c}
\hline
Dataset & Expression\\
\hline
\multirow{5}{*}{US} & $c_3\ln{\left(c_{1} w_i + w_j^{3} w_i \left(\frac{c_{2} c_{3}}{c_{2} + d}\right)^{c_{4}} \right)}$\\
&$c_{1} + c_4 \left(c_{3} d w_i\right)^{- \left(\left(c_{5} c_{6} w_j\right)^{c_{6}}\right)^{- d}} \left(c_{2} + c_{3}^{w_i}\right)$\\
&$c_{1} \left(c_{3} + d\right)^{- c_{2}} \ln{\left(c_{2} + w_j \right)} + \left(w_i \left(c_{4} + w_j\right)\right)^{c_{5}}$\\
&$\left(w_iw_j\right)^{c_{1}} e^{\left(c_{1} d\right)^{c_{2}} \left(c_{3} + d\right)}$\\
&$c_{1} c_{1}^{- \tanh{\left(\frac{c_{3} + d}{d} \right)}} w_i^{c_{2}}w_j^{c_{2}}$\\
\hline
\multirow{5}{*}{England} & $c_{1}^{2 d \ln w_i } \left(c_{2} w_j + c_{3}\right)/(c_{2} + d + w_j)$\\
&$c_{1} e^{\frac{c_{2} d}{c_{3}/d + d + w_i^{c_{4}} \left(c_{2} + w_j\right)}}$\\
&$c_{1} \left(e^{\frac{c_{2} \left(d + w_i\right)}{c_{3} + w_j + w_i} + c_{4}} + \ln{\left(w_iw_j \right)}\right)$\\
&$c_{1} \tanh{\left(\frac{c_{2} \left(c_{3}^{d} w_j w_i\right)^{c_{2}} \left(c_{4} + w_j\right)}{d} \right)}$\\
&$c_{1} \left(c_{2} w_iw_j \right)^{c_{1}} + c_{3}^{\frac{d}{c_{4} + w_j}} \left(c_{2} + c_{5} w_i\right)$\\
\hline
\end{tabular} 
\label{tab:bms}
\end{table}

\end{document}